\documentclass{ieeeaccess}
\usepackage{cite}
\usepackage{amsmath,amssymb,amsfonts}
\usepackage{algorithm}
\usepackage[end]{algpseudocode}
\usepackage{graphicx}
\usepackage[table,xcdraw]{xcolor}
\usepackage{textcomp}
\usepackage{cite}
\usepackage{amsmath,amssymb,amsfonts}
\DeclareMathOperator*{\argmax}{arg\,max}
\DeclareMathOperator*{\argmin}{arg\,min}
\usepackage{graphicx}
\usepackage{textcomp}
\usepackage{xcolor}
\usepackage{tabularx}
\usepackage{threeparttable}
\usepackage{array}
\newcolumntype{L}{>{\raggedright\arraybackslash}X}
\usepackage{multirow,multicol}
\usepackage{caption}
\usepackage[binary-units]{siunitx}
\usepackage[super]{nth}
\usepackage{adjustbox}
\newcommand{\tabitem}{~~\llap{\textbullet}~~}
\DeclareSIUnit\decibelm{dBm}
\DeclareSIUnit\packets{packets}
\DeclareSIUnit\pct{percentile}
\DeclareSIUnit[per-mode=symbol]\bps{\bit \per \second}
\DeclareSIUnit[per-mode=symbol]\kbps{\kilo\bps}
\DeclareSIUnit[per-mode=symbol]\Mbps{\mega\bps}
\def\BibTeX{{\rm B\kern-.05em{\sc i\kern-.025em b}\kern-.08em
    T\kern-.1667em\lower.7ex\hbox{E}\kern-.125emX}}
\begin{document}
\history{Date of publication xxxx 00, 0000, date of current version xxxx 00, 0000.}
\doi{10.1109/ACCESS.2017.DOI}

\title{Signaling Design for Cooperative Resource Allocation and its Impact to Reliability}
\author{\uppercase{Rasmus Liborius Bruun}\authorrefmark{1},
\uppercase{C. Santiago Morejón García\authorrefmark{1},  Troels B. Sørensen\authorrefmark{1}, Nuno K. Pratas\authorrefmark{2}, Tatiana Kozlova Madsen\authorrefmark{1} and Preben Mogensen\authorrefmark{1}\authorrefmark{2}}}
\address[1]{Wireless Communication Networks Section, Department of Electronic Systems, Aalborg University, Denmark}
\address[2]{Nokia Standardization, Aalborg, Denmark}

%\tfootnote{This paragraph of the first footnote will contain support 
%information, including sponsor and financial support acknowledgment. For 
%example, ``This work was supported in part by the U.S. Department of 
%Commerce under Grant BS123456.''}

\markboth
{Author \headeretal: Preparation of Papers for IEEE TRANSACTIONS and JOURNALS}
{Author \headeretal: Preparation of Papers for IEEE TRANSACTIONS and JOURNALS}

\corresp{Corresponding author: Rasmus L. Bruun (e-mail: rlb@es.aau.dk).}

\begin{abstract}
Decentralized cooperative resource allocation schemes for robotic
swarms are essential to enable high reliability in high throughput
data exchanges. These cooperative schemes require control
signaling with the aim to avoid half-duplex problems at the
receiver and mitigate interference. We propose
two cooperative resource allocation schemes, device sequential
and group scheduling, and introduce a control
signaling design. We observe that failure in the reception of
these control signals leads to non-cooperative behavior and to
significant performance degradation. The cause of these failures are identified and specific countermeasures are proposed and evaluated. We compare the proposed
resource allocation schemes against the NR sidelink mode 2
resource allocation and show that despite signaling has
an important impact on the resource allocation performance,
our proposed device sequential and group scheduling resource
allocation schemes improve reliability by an order of magnitude
compared to sidelink mode 2.

\end{abstract}


\titlepgskip=-15pt

\maketitle

%%%%%%%%%%%%%%%%%%%%%%%%%%%%%%%
%%%%%%%%%%%%%%%%%%%%%%%%%%%%%%%
%%%                         %%%
%%%         SECTION         %%%
%%%                         %%%
%%%%%%%%%%%%%%%%%%%%%%%%%%%%%%%
%%%%%%%%%%%%%%%%%%%%%%%%%%%%%%%
% !TeX spellcheck = en_US
\section{Introduction} \label{sec:intro}

% Broad vision of swarms and flocks, general applicability and advatageus 
%   Requirements for the envisioned usecases - throughput,latency, organization (decentralized)
% The wireless context - decentralized, selforganized, ... -> role (enabler). Touch on standards - 802.11p bd 802.11s 802.15.4 (IEEE), Bluetooth (SIG), Cellular (3GPP)
% Mode~2 -> what can it do (how it works) and what are the limitations
% Potential extensions
% Other decentralized resource allocation schemes and algorithms.
% Our preliminary (published) work in PIMRC - 2 schemes. They were not evaluated with realistic signaling
% contribution: Same as conference 2

The density of connected devices is growing rapidly. %Communication ability is given to an increasing amount of things, as it is becoming feasible to connect even the light bulbs we have a home. 
Nowadays it is insufficient to have connectivity only in smartphones. Wireless connectivity is expanding to wearables, domotics, automotives, etc., to make our lives simpler, safer and more convenient. As connectivity becomes omnipresent, the basis for a new form of collaboration has been created. Nature has inspired many technological leaps, and the collaboration of simple entities is a well known phenomenon in the animal kingdom, where ants, birds, bees, fish and a plethora of other species have learned to benefit from collaboration, allowing them to unite efforts and enable them to achieve complex tasks. The behavior of swarming, flocking and schooling serves as inspiration for the collaboration which has become possible between connected electronic devices. The first use cases have already been envisioned, e.g:
\begin{itemize}
\item In manufacturing, swarms are envisioned to enhance production lines, by enhanced flexibility and adaptability enabled by better communication \cite{rodriguez_5g_2021}.
\item In search and rescue flocks, drones are envisioned to cover land quickly and with short response time, thus vastly cutting the critical time to find lost persons in the debree of a collapsed building, people lost at sea, in a forest, etc. In such operations it is life critical to locate the missing persons as soon as possible \cite{arnold_uav_2020}.
\item Within the agricultural industry, in \cite{agriculture} a monitoring and mapping system guides autonomous weeding robots. This system maps the field by using a swarm of UAVs to patrol it. The system provides weed's presence identification and location of different intervention urgency areas.
\item In domotics (smart home and office) the collection of connected smart devices (each with a distinct sensing, actuating or service function) will collaborate like the bee swarm maintaining the hive, to efficiently monitor the state of the building and provide an optimal indoor environment while minimizing the energy bill cost \cite{jordehi_optimal_2019,malik_optimal_2020}. The robot vacuum will operate where needed, but at the most convenient times and the heating, ventilation, and air-conditioning (HVAC) will be adjusted ad-hoc to provide the perfect indoor climate at all times of day and year \cite{wick_towards_2017,malik_optimal_2020}.
\item In automotive, connected devices will be vital to maintain a streamlined transportation infrastructure where the transportation needs of humans, and goods can be met in the safest and most seamless way possible. The sensors around the transportation grid will provide real-time traffic updates to the active vehicles which will negotiate their optimal routes as they drive in dense platoons not unlike ants and migrating birds \cite{hedge_enhanced_2019,wu_performance_2020}.
\end{itemize}

These are just to mention a few of the present use cases. Undoubtedly, the most revolutionary applications of swarm robotics have yet to be discovered as technologies mature and become accessible. %There is a continuous stride towards higher throughput, lower latency and better reliability. As communication performance improve, the possibilities of swarm communication increase. % The chicken and the egg...
Common for the aforementioned use cases and the nature of swarms is the need for communication between devices within proximity. In theory, direct one hop communication between devices has the shortest possible latency and best utilization of time-frequency resources. Also, it provides good conditions to obtain high reliability which we define as the probability that a receiver successfully decodes a received message within an application's latency requirement. However, achieving these benefits will require smarter solutions. For that reason our efforts are concentrated on decentralized communication where all devices engage in communication on equal terms and no coordination from network or one specific device is needed for communications to take place. Additionally we are concerned with pushing beyond the current state of the art, thus focus on how to improve throughput and reliability at reduced latency.

\subsection{Decentralized Wireless communication}
Different solutions exist for decentralized wireless communications, however to achieve adoption and wide spread usage, standardization is indispensable. Standardized wireless communication technologies enable different manufacturers to produce compatible products. This aids competition and will result in larger supply of products at lower cost. The most known standards are governed by IEEE, Bluetooth SIG and 3GPP. 

% PAN
Bluetooth SIG governs the Bluetooth standard which is a personal area network technology. Bluetooth Classic refers to the original Bluetooth protocol stack which was originally meant as a wireless alternative to a cabled connection, e.g. between headset and phone. In version 4.0 of the Bluetooth Core Specification, the Bluetooth Low Energy protocol stack was introduced. Bluetooh Low Energy is incompatible with Bluetooh Classic and designed for low power consumption. Both Bluetooth stacks operate in the unlicensed 2.4 GHz Industrial, Scientific and Medical (ISM) band. The Bluetooh Mesh specification \cite{mesh_working_group_mesh_2019} was adopted in 2017 to allow Bluetooth technology to cater applications which include multiple device networks. %not centered around a the start topology of the pico nets
% LAN

The IEEE 802.15.4 standard is a low-data-rate, low cost and low power physical and MAC layer specification \cite{802_15_4}. It was originally conceived to enable low cost personal area networks between ad-hoc devices, and operates in the ISM bands between 0.8 and 2.4 GHz. IEEE also governs the 802.11 standard, which is a specification of protocols for wireless local area networks. The amendments 802.11a/b/g/n/ac/ax refer to WiFi networks, which connect computers and smartphones to the internet via an access point. However, 802.11s, 802.11p and 802.11bd are amendments directed at device to device applications. The 802.11s amendment enables mesh networking in which packets are routed according to one of the supported protocols. Dedicated short-range communication is supported by the 802.11p and the upcoming 802.11bd amendments. The target of these amendments is to enable vehicular communication in the 5.9 GHz Intelligent Transport Systems (ITS) band.

%delimitation> We are not considering IEEE and bluetooth standards
The main challenge that Bluetooth SIG and IEEE governed standards have is their operation in the unlicensed spectrum bands where they need to abide by either listen before talk or duty cycle restrictions \cite{cept_erc_2021,etsi_electromagnetic_2012}. For this reason, these standards are vulnerable to interference and low spectral efficiency which limits their achievable throughput and latency performance \cite{d2d_beyond4G}.

% Something about the Citizen Broadband Radio Service (CBRS) band of 150MHz at 3.5 GHz in the US. Which was released in 2020 for 5G use without obtaining a lincense. Could be inistalled in a per-building scenario (though it seems the users will need to pay to use it). It is thought to be a better option for airports, fatories and ports for connectivity and ideeal for 5G services. 
% https://www.federatedwireless.com/wp-content/uploads/2017/09/Mobile-Experts-CBRS-Overview.pdf
% https://en.wikipedia.org/wiki/Citizens_Broadband_Radio_Service
In the United States, the 3.5 GHz Citizen Broadband Radio Service (CBRS) band with a bandwidth of 150 MHz was established in 2015 to allow shared commercial usage in the band \cite{mun_cbrs_2017}. Up to 70 MHz is licensed by census tract (limited geographical region) allowing factories, airports and the like to license the band and utilize it for a dedicated network. This licensing arrangement is interesting for future use cases of e.g. cellular technologies which already operate in this band in other parts of the world.

3GPP standardizes cellular communication. The concept of device-to-device communications appeared within 3GPP release 12, with the development of proximity services (ProSe). The most recent version of the standard is release 16. Among other things, it includes decentralized device-to-device communications in the form of NR sidelink resource allocation mode~2 (mode~2). The mode~2 resource allocation is explained in detail in section \ref{sec:mode2}. The main performance constraints of mode~2 are caused by the presence of half-duplex problems and multi-user interference \cite{sidelink_drawbacks}. Half-duplex refers to the limitation a transceiver has, since it is not able to receive and transmit simultaneously. The problem arise when two communicating transceivers transmit to each other simultaneously rendering both unable to receive. To overcome these issues, inter-UE coordination is being discussed for the upcoming release 17 \cite{collaborativeRA_3gpp}. Here the concept of cooperation/coordination is adopted as an option to be added on top of mode~2 which should aid in mitigating half-duplex and interference problems. Two coordination schemes are agreed upon: Inter-UE coordination scheme 1 and scheme 2. In the former, upon request, the receiving UE-A assists the transmitting UE-B in resource allocation by indicating a set of preferred/non-preferred resources for the transmitting UE-B; the latter allows the receiving UE-A to notify the transmitter that the resource selected by the transmitter results in expected/potential and/or detected conflicts.

%As mode~2 operates in the licensed spectrum it is not constrained by listen before talk regulations, but the current state of mode~2 has shortcomings which are currently being addressed by coordination techniques. 
The inter-UE coordination framework being introduced in 3GPP Rel.17 does not target  swarm use cases where a group of UEs have to exchange information. %More specifically, it focuses on ensuring that a single UE-B is able to perform reliable transmission (i.e. avoid half-duplex and interference) via the coordination from a UE-A. The signaling to achieve this is based on the UE-B making a inter-UE coordination request, and then the UE-A providing the inter-UE coordination information (e.g. in the form of preferred and/or non-preferred resources). 
In other words, the signaling is pair based and not efficient for use cases where a group of UEs requires coordination information.

The scope of this paper is to introduce and evaluate a cooperative inter-UE coordination scheme suitable for group coordination.
%There is therefore the need to introduce inter-UE coordination for groups, which is the scope of this paper.
%The former consists of sending coordination information from the receiver UE to its transmitter UE. This coordination information contains the set of resources preferred/or not preferred for the data transmission. The latter consists of the receiving UE sending information to the transmitter about the presence of expected/potential and/or detected resource conflicts on the resources indicated by the transmitter UE.% in its sidelink control information (SCI).  

\subsection{Cooperative communications} \label{subsec:coopComm}
Consensus on the use of time-frequency resources is the basis of multi-user communications. In decentralized communication systems one way to achieve high throughput, high reliability and low latency is to reach consensus in the usage of time-frequency resources via cooperative resource allocation. Authors in \cite{consensusRA} introduced two consensus communication protocols, the first a gossip-based (multi-hop message diffusion) and the second a broadcast (single-hop message diffusion) communication protocol. In both protocols, a set of UEs (validators) validates and commits the proposed action (vacant frequency band) made by the proposer UE. The consensus protocols have low latency and high reliability that could support mission-critical and real-time tasks as long as consensus decisions change infrequently. The validation process may take some time due to the number of validators, and conversely if this number reduces, reliability may suffer. Therefore, there is a need for a balance between reliability and latency. The main advantage of the consensus algorithm is its resilience to UEs with malicious intent. 

In systems without "malicious UEs", the consensus procedure is no longer necessary since it is assumed that all nodes will follow the specified resource allocation procedure. Consequently, an optimal resource allocation scheme can be reached faster. Authors in \cite{swarm_algorithms_distributed_RA} developed resource allocation algorithms inspired by a bio-swarming behavior. The presented methods rely on multiple iterations before they converge to an optimal resource allocation. In \cite{distributed_ra_quadratic_time} the authors present a distributed resource allocation scheme which converges in quadratic time. The convergence is dependent on the number of devices, because each device is involved in execution of the algorithm.

%A distributed resource allocation on dynamic networks in quadratic time is presented in \cite{distributed_ra_quadratic_time}. A transmitter device broadcasts a derivative and Lipschitz constant (used for optimization) to its respective neighbors. Neighbors are identified with the smallest derivative by sending them a message containing the delta-calculation. Devices receiving delta-calculation send an acceptance to the transmitter with the largest value and reject the rest of them. The time this algorithm takes to converge is relative to the number of devices in the network and could take thousands of iterations to complete. 

\begin{figure}[t!]
	\centerline{\includegraphics[width=\linewidth]{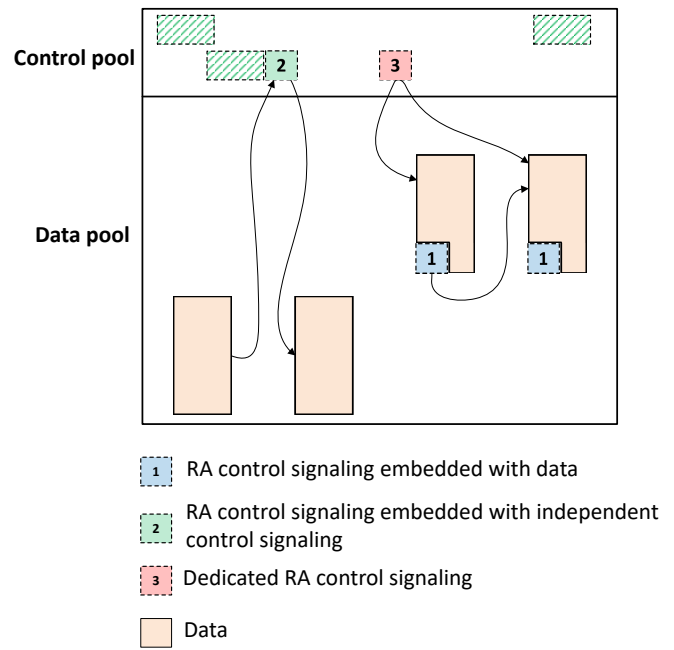}}
	\caption{Methods to exchange signaling: i) Embedded with data (blue), ii) Embedded with independent signaling (green), and iii) Dedicated (red)} 
	\label{fig:sigCategories}
\end{figure}

Although these distributed consensus and resource allocation schemes achieve full alignment of the swarm members and an optimal resource allocation, the involvement of the majority of the swarm members in the allocation process is detrimental to the latency as the swarm size grows. Instead, it is desired that changes to the resource allocation can be performed locally among one or several sub-sets of swarm members, such that overhead in the form of control signals is limited. Mode~2 operates like this (a more detailed explanation is provided in Section \ref{sec:mode2}), where control signals are embedded with the data transmissions and thus only reach nearby swarm members (illustrated by control signal method 1 in Fig.~\ref{fig:sigCategories}).

%Consensus and distributed resource allocation schemes rely on involvement of most devices in the swarm, thus even the low latency schemes like \cite{consensusRA} will incur additional latency as the swarm grows. The resource allocation schemes with the objective to optimize the resource allocation require multiple rounds of communication, where each participant need to transmit at least once every round. This is also detrimental to the latency. 

%Thanks to wireless communication standards and processes to certify compliance as well as legislation on the usage of the radio spectrum, real world communication devices are not limited by the consensus among devices, but rather the limits imposed by standards and legislation. This is advantageous for latency, because it allows schemes where only a subset or even a single device may acquire a resource allocation without having to notify every other node in the network.

% This is about signals required for coordination
%In that light, the 3GPP standard is a solid basis for resource allocation. Some of its shortcomings have been addressed in recent work. When working with existing standards, an important point is to consider how to include the coordination information. Fig.~\ref{fig:sigCategories} illustrate three methods to include additional resource allocation signals in the communication.

In recent literature, cooperative extensions to the existing NR mode 2 standard have been suggested in an effort to address shortcomings in the current version. The continuous collision problem of the semi-persistently scheduled (SPS) transmissions is tackled in \cite{peng_age_2020} by allowing a third UE to piggyback with its own transmission an indication that continuous collision is (likely) taking place in another resource (method 1 in Fig.~\ref{fig:sigCategories}). It is a reactive scheme where collisions are resolved rather than avoided. The scheme depends on other UEs being able to assist.

In \cite{bonjorn_cooperative_2019} authors introduce a counter in the SPS signaling indicating the time of reselection, i.e. as method 1 in Fig.~\ref{fig:sigCategories}. Within each SPS transmission, a procedure is proposed to adjust the counters such that no UE will be reselecting in the same transmission time interval, thus mitigating SPS collisions. The procedure is proactive as it tries to mitigate future collisions, but the procedure is designed for low density swarms. Additionally, failures on the initial transmission are not handled.

The piggybacking of control signals to periodic safety messages in \cite{jeon_reducing_2018} indicates the future resource allocation of SPS transmissions (method 2 in Fig.~\ref{fig:sigCategories}). The next SPS allocation is performed before the end of the current SPS transmission, allowing time to reselect the SPS in case of potential conflicts.

With the aim of addressing the mode 2 limitations (i.e. susceptibility to half-duplex and interference), in \cite{cooperative_swarm_comm} we introduced two cooperative resource allocation schemes, device sequential and group scheduling, which follow the framework of 3GPP but have different resource allocation algorithms and cooperation schemes. The coordination scheme of both the device sequential and group scheduling approach allow coordination where message exchange is only required for devices in proximity and with immediate need for communication resources. The additional signaling required could partly be piggybacked to existing discovery messages (method 2 in Fig.~\ref{fig:sigCategories}) and by the introduction of a dedicated control signal (method 3 in Fig.~\ref{fig:sigCategories}). Under assumption of perfect exchange of control messages, the proposed schemes far outperformed mode~2 \cite{cooperative_swarm_comm}.

In this article we recapitulate the device sequential and group scheduling resource allocation schemes and additionally design the required signaling exchange to evaluate its impact on data reception reliability. We show how device sequential and group scheduling schemes provide significant performance improvement over the baseline despite introducing signaling overhead. Our evaluation focuses on the causes of data reception failures, thereby allowing a deeper analysis of the signaling design, its impact on the resource allocation, and the resulting data reception performance. Additionally, it provides us information to propose techniques to overcome such failures and evaluate their impact on the final data reception reliability. The specific contributions in this article are:
\begin{itemize}
	\item [$-$] Signaling design to enable distributed cooperation for the proposed device sequential and group scheduling cooperative resource allocation schemes.
	\item [$-$] Evaluation of signaling overhead on the performance.
	\item [$-$] A methodology of analysis for separating communication failures and identifying the most impacting causes, thereby deepening the understanding of performance differences and focal points for further enhancements.
	\item [$-$] Techniques to enhance the signaling reliability and overall swarm application performance, based on the specific failure causes. 
	%Analysis of the reception failure causes. This provides grounds for two mechanisms which will improve the overall reception failure probability. In addition, this analysis methodology will give insight as to why the proposed cooperative resource allocation schemes outperform the existing 5G NR sidelink mode~2.
\end{itemize}

% What will be presented in this paper?
We continue with Section \ref{sec:sysmdl} presenting the assumptions, notation and the baseline mode 2 allocation scheme. In Section \ref{signDesign} we present the cooperative resource allocation schemes device sequential and group scheduling. Control signaling design for the cooperative schemes are presented in Section  \ref{sec:cntrlSig}. Section \ref{sec:sim} outlines the simulation setup and the simulation results and enhancement techniques are presented and evaluated in Section \ref{sec:eval}. Concluding remarks are made in Section \ref{sec:con}.

\section{System Model and Notation} \label{sec:sysmdl}
% - Here we'll provide the system model and framework for our research, and define the objective function and optimization problem.
Consider a system  of $N$ UEs engaging in proximity communication, enabled by their omnidirectional antennas and half-duplex radios. At any point in time, a UE is either not involved in proximity communication, and therefore not transmitting data messages, or the UE is involved in proximity communication, defined by UEs being within (a device-centric) critical communication range of $r_c$. We differentiate between \emph{data messages}, defined as the information bits transmitted for the purpose of some swarm application, and \emph{control signals}, defined as the transmitted bits which serves a supporting function not directly related to the swarm application. The proximity communication consists of transmitting and receiving multi-casted data messages of size $x_d$ bytes with a $d_p$ seconds periodicity to and from all UEs within proximity; i.e., a UE will transmit data at a rate of $t_d = x_d/d_p$ bytes per second during proximity communication. The need to transmit data messages is determined based on proximity: the \emph{ready time} is the moment in time when a data message is ready from application layer. A maximum latency of $l$ seconds can be tolerated from the ready time until the message is delivered at all intended destinations. Combined, the ready time and latency budget defines the deadline of the data messages. The data message becomes useless after the deadline and will be discarded. Some control signals might be exchanged regardless of proximity.

%- ues
%- type of transmissino
%  -focus on data
%- transmission resources, time ,frequency, 5G framework
%- objective function

We follow the 3GPP system framework \cite{PHY_3gpp} where communication is based on Orthogonal Frequency Division Multiplexing (OFDM) on a frequency band of bandwidth $B$. The frequency resource is shared between UEs by time division multiple access (TDMA). The smallest allocation unit is called a slot and has duration $d_s$, which is configurable based on the selected numerology. For simplicity we adhere to numerology 2, the highest numerology available for frequency range 1, which results in the shortest slot duration. We refer to time slots by their index $s$ in the set $\mathcal{S} = \{1,2,...S\}$, which spans the lifetime of the network. For simplicity we assume UEs to have the same transmission requirements and be time synchronized, i.e. following the 5G NR procedure explained in \cite{lien_3gpp_2020}. In the following sections we use the notation in Table \ref{tab:not}.

\begin{table}[]
	\caption{Notation}
	\label{tab:not}
	\resizebox{0.5\textwidth}{!}{%
		\begin{tabular}{ll}
			\hline
			Symbol                               	   & Meaning                                          			\\ \hline
			$N$               					   & Total number of autonomous robots                			\\
			$W$               					   & Bandwidth for data transmissions               			\\
			$S$			                           & Number of slots in the lifetime of the network             \\
			$r_e$			                           & Extended cooperation range                       			\\
			$r_c$			                           & Critical cooperation range                       			\\ 
			$n_s$			                           & Number of slots requested by a UE for its transmission     \\ 
			$\mathcal{N}$			                   & Set of UE IDs $\mathcal{N} = \{1,2,...,N\}$				 \\
			$\mathcal{S}$			                   & Set of time slots 	$\mathcal{S} = \{1,2,...,S\}$			 \\
			$\mathcal{C}$			               	   & Set of indices of candidate slots for possible resource allocation \\
			& $\mathcal{C}=\{1,2,...,40\}$ indicates the slots with indices 1 to 40  \\ 
			$\mathcal{R}$                 	           & Set of information about utilization of slots in $\mathcal{C}$       \\ 
			$\mathcal{R}_s$			                   & Resource occupancy determined by sensing procedure         \\ 
			$\mathcal{R}_e$			                   & Resource occupancy determined by exchange of control signals\\ 
			$\mathcal{A}$			                               & Slots allocated for requested transmission(s)              \\ 
			$s$			                               & Indicates a unique slot in $\mathcal{S}$                        		\\
			$o$			                               & Indication of the occupancy in a slot   \\
			$d_p$			                           & Transmission periodicity                       			\\
			$x_d$			                           & Size of data message in bytes                       			\\
			$t_d$			                           & Data message data rate                       			\\
			$p_{tx}$			                       & Transmission power\\
			$T$			                           	   & Thermal noise power                      			\\
			$g_{n,n'}^s$			                   & Channel gain between UEs $n$ and $n'$ in slot $s$		\\
			$\gamma_{n,n'}^s$			               & SINR on transmission between UEs $n$ and $n'$ in slot $s$    	\\
			&                        \\ \hline
		\end{tabular}%
	}
\end{table}

%\subsection{Problem formulation} % TODO
%% Objective function
%The system consists of $N$ UEs sharing the transmission resources consisting of a bandwidth $B$ separated into slots . 
When a UE with id $n \in \mathcal{N} = \{1,2,...N\}$ generates data in a slot $s$, this data is associated with a group of receivers $\mathcal{N}' \subset \mathcal{N}$ where $n' \in \mathcal{N}' : n\neq n', \mathrm{dist}(n,n') < r_c$. The function $\mathrm{dist}(n,n')$ returns the euclidean distance between UEs $n$ and $n'$. 
For simplicity, we assume every transmission is subject to the same transmission power $p_{tx}$. The channel gain on transmission from $n$ to $n'$ in slot  $s$ is given as $g_{n,n'}^s$ and the gain from interfering transmissions is given as $g_{k,n'}$  $\{k : k \in \mathcal{K} \subset \mathcal{N}, k \neq n, k\neq n'\}$. The channel gains are modeled as the combined effect of path loss and shadowing, where the shadowing component on different links is correlated. 

When a slot is used for transmission, the SINR on a link between $n$ and $n'$ in slot $s$ is calculated according to
\begin{equation}
	\gamma_{n,n'}^s = \frac{p_{tx} g_{n,n'}^s}{T+\Sigma_{k \in \mathcal{K}}p_{tx}g_{k,n'}^s} 
\end{equation}
where $T$ is the thermal noise power.

Based on the ready times and latency requirement, the 3-dimensional data transmission matrix can be obtained
${\mathbf{D}}_{N \times N \times S} = [\delta_{n,n',s}]$

\begin{equation}
	    \delta_{n,n',s} = 
	\begin{cases}
		s+\lfloor\frac{l}{d_s}\rfloor, 			& \parbox[t]{5.3cm}{ if  $n$  generates data in slot $s$  which should be transmitted to $n'$  within latency $l$} \\
		0,              & \text{otherwise}
	\end{cases}
\end{equation}

The problem is to determine an allocation, indicated by the allocation matrix ${\mathbf{A}}_{N \times S} = [\alpha_{n,s}]$ where the maximum number of UEs can be supported in the swarm.

\begin{equation}
	\alpha_{n,s} = 
	\begin{cases}
		1, 			& \parbox[t]{3.3cm}{if $n$ transmits in slot $s$ }\\
		0,          & \text{otherwise}
	\end{cases}
\end{equation}

such that for each nonzero entry $\delta_{n,n',s}$ in ${\mathbf{D}}$, the corresponding transmissions can be determined as the nonzero entries of ${\mathbf{A}}$ in the corresponding row $n$ and the columns in the interval $[s;s+\lfloor\frac{l}{d_s}\rfloor]$. We refer to this interval as the allocation interval, as it is the slot interval in which a UE can be allocated transmission resources for a given data packet. Let $\Delta_{\delta_{n,n',s}} = \{\alpha_{n,r} : r \in S, s <= r <= s+l\}$ be the set of slots $n$ utilize for the transmission of data $\delta_{n,n',s}$ to $n'$. The combined SINR of the transmissions relating to the same data message is calculated as 

\begin{equation}
	\gamma_{\delta_{n,n',s}} = 2^{\frac{1}{K} \sum_{r \in \Delta_{\delta_{n,n',s}}} {log}_{2}(1+\gamma_{n,n'}^r)}-1
\end{equation}
which is also known as the mean instantaneous capacity method used to determine an effective SINR mapping. Thus a set can be defined as $\Gamma = \{ \gamma_{\delta_{n,n',s}} : \delta_{n,n',s} \neq 0, n \in N, n' \in N', s \in S \}$ and the optimization problem is formulated as

\begin{subequations}
	\begin{alignat}{2}
		&\!\argmax_{\mathbf{A}_{N \times S}}    &\qquad& N\\
		&\text{subject to} &      & \frac{1}{|\Gamma|}\Sigma_{\gamma_i \in \Gamma} \mathrm{bler}(\gamma_i) < f_p  ,\\
		&                  &      & \Sigma_{i \in N'}\alpha_{i,s} <= 1  \label{eq:constraint2}
	\end{alignat}
\end{subequations}

where $\mathrm{bler}(x)$ is a mapping function which maps a certain SINR to a block error rate, following the physical layer abstraction given in \cite{lagen_new_2020}. The first constraint guarantees that the system failure probability does not exceed a required failure probability requirement, $f_p$. The second constraint ensures that no two UEs within critical cooperation range transmit simultaneously, thereby avoiding half-duplex problems.

The problem of determining the allocation matrix $\mathbf{A}$ (like the problem formulated in \cite{zhang_interference_2013}) is NP-hard, thus no algorithm can be found to determine the optimal solution within polynomial time. Additionally, due to the potential overlap of allocation intervals of different UEs, in search of the optimal solution, the entire lifetime of the network should be considered. Therefore, it is not feasible to find an optimal solution to this problem, and instead we deal with heuristic methods to efficiently determine suboptimal solutions to the allocation problem in a decentralized manner. We note that these approaches limit the scope of each round of allocation such that only a subset of slots in $\mathcal{S}$ is considered. Furthermore, due to the decentralization of the system, the allocation decision can be delegated to each UE, which might have a limited knowledge about the allocation decisions of other UEs. We will see that knowledge of other UEs allocation decisions is important for the performance of the system, as it might help avoid half-duplex allocations and reduce interference. In the next section, the state of the art decentralized resource allocation algorithm of 5G NR is presented.

\subsection{Baseline resource allocation scheme (Mode~2)} \label{sec:mode2} \label{sensingSL}
On the sidelink, UEs can transmit directly to each other by performing 5G NR sidelink resource allocation mode 2. Mode~2 \cite{3gpp_ts_2020} relies on the signaling exchanged in the sidelink control information (SCI). The SCI is transmitted as part of a data message as depicted by control signal method 1 in Fig.~\ref{fig:sigCategories}. The SCI carries information which is necessary for decoding of the data transmission, but more importantly (in a resource allocation perspective) it indicates the periodicity of the transmission, i.e. the future resources reserved for this semi-persistently scheduled (SPS) transmission. SPS transmissions reduce the overhead of resource allocation by introducing predictability, which allow other UEs to avoid allocation of conflicting resources. In addition, the UEs can reuse the resource allocation of one data message of subsequent data messages. This is a key concept of mode~2 by which UEs autonomously allocate resources. For completeness, we summarize the two stages of mode 2 below. 

\subsubsection{Sensing stage}
Sensing is performed on a \emph{sensing window} which spans the bandwidth configured for mode~2 transmissions and a time span no longer than 1 s leading up to the selection stage. The goal of the sensing is to determine a set of candidate resources. Initially a set of candidate resources of size $M_{total}$ is defined. Resources are removed from the candidate set if a SCI received during the sensing window indicates that the candidate resource is reserved by another UE \textbf{and} the measured reference signal received power (RSRP) on the SCI is above a threshold. If the resulting candidate set is smaller than 20\% of $M_{total}$, the threshold is increased by 3 dB and the discarded candidate resources are re-evaluated, i.e. re-introduced to the candidate set if the RSRP is below the threshold.

\subsubsection{Selection stage} \label{sec:resourceSelection}
In the selection stage the resource allocation algorithm is performed. It consists of selecting the requested resource(s) randomly among the candidate resources. If the resources are reoccurring with a given periodicity, the SPS re-selection counter is initialized \cite{3gpp_ts_reselection}. At each transmission using the allocated  resource, the counter is decremented. Once the counter reaches zero, re-selection is performed according to the mode~2 resource allocation. \\

%Congestion control subdues transmissions based configured priority levels. ---To check if it is necessary to add this

% Resource allocation
% Coordination (lack hereof)
%The coordination scheme of mode~2 is limited to the control signal in the SCI indicating the future resource usage, which is used in the sensing procedure. 
%\subsubsection{title}

\noindent The advantage of mode~2 is the autonomy of the procedure. It is only affected by the information it is able to obtain during the sensing window, and the delay introduced by determining the candidate slots is fixed. The disadvantage is that the simple coordination might cause two UEs with close ready times to allocate overlapping resources, resulting in half duplex problems. Additionally, the random nature of the allocation can cause sub-optimal performance.

%Next, we introduce our cooperative resource allocation schemes.
%Following, we introduce our cooperative resource allocation schemes. Both were built under mode 2 (i.e., 3GPP sidelink design). Even though random resource allocation and partial (i.e., mode 2 sensing, where the sensing period is punctured or reduced in time) are used by other methods, they are not compatible with 3GPP sidelink design. Therefore, a comparison between our proposed schemes and other resource allocation methods is not feasible.  

Following, we introduce our cooperative resource allocation schemes. Both were built to comply with the 3GPP sidelink framework and its possible extensions. The sidelink framework is different from the framework of ISM-band technologies, where listen-before-talk and duty cycle restrictions are essential bounds on the resource allocation. Therefore, mode 2 acts as the baseline to which we compare our proposed allocation schemes.

\section{Proposed cooperative resource allocation schemes} \label{signDesign}
%In this section we present the three resource allocation schemes, which will be the focal point of our study.
The cooperation scheme refers to the distribution of the resource allocation and related functions. It answers the question of who will perform the resource allocation, when, and based on what information. As in Fig.~\ref{fig:sigCategories}, we assume the bandwidth is divided into two adjacent frequency resource pools to accommodate control transmissions and data transmissions, respectively. The UEs are able to either transmit or receive in both resource pools simultaneously, but due to the half-duplex constraint, simultaneous reception and transmission is not possible. Section \ref{sec:cntrlSig} will explain what types of signals are transmitted in the control pool. The control pool signals are intended for UEs within extended communication range of $r_e$ m. %For convenience, the notation is summarized in Table \ref{tab:not}.

\begin{algorithm}[t]
	\caption{Resource allocation}
	\label{alg:RA}
	\textbf{Input:} $\{(n_s,\mathcal{R} = \{\mathcal{R}_s \cup \mathcal{R}_e\},\mathcal{C})_k\}, k = 1,2,...,{K}$\\
	\textbf{Algorithm}:	
	\begin{algorithmic}[1]		
		\For {each k in descending order of number of UEs within \emph{critical cooperation range}}
		\State $\mathcal{P} = \{(s,o)_i \in \mathcal{R}_k, o_i \neq \infty,s_i \ in \mathcal{C}_k, i = 1...|\mathcal{R}_k|\}$
		%\State $P \subseteq R : P = \{s,o\}_i \wedge o_i \neq \infty)$ % P is the subset of R where o_i is not infinity
		%\State Select from P the $n{s,k}$ pairs with the lowest $o_i$
		\For {$n = 1, 2, ..., n_{s,k}$}
		\State \parbox[t]{\dimexpr\linewidth-\leftmargin-\labelsep-\labelwidth}{%
			$\argmin_i o_i \in P $}
		\State \parbox[t]{\dimexpr\linewidth-\leftmargin-\labelsep-\labelwidth}{%
			$\mathcal{A}_{k} \leftarrow \mathcal{A}_{k} \cup s_i$}
		\State \parbox[t]{\dimexpr\linewidth-\leftmargin-\labelsep-\labelwidth}{%
			$\mathcal{P} \leftarrow \mathcal{P} - (s,o)_i$}
%		\begin{spacing}{0.60}
%			\begin{singlespace}
%			\end{singlespace}
%		\end{spacing}
		\EndFor
		\EndFor
	\end{algorithmic}
	\textbf{Output:} $\{\mathcal{A}\}_k$
\end{algorithm}

% Both schemes share resource allocation algorithm, they just coordinate differently
Both proposed resource allocation schemes described here share the same basic resource allocation algorithm (Algorithm)~\ref{alg:RA}). We differentiate between the \emph{allocating} UE which is the UE executing the resource allocation algorithm, and the \emph{requesting} UE(s), which is the UE(s) requesting an allocation from the allocating UE. The allocating and requesting UE can be the same UE. The input is the tuple $(n_s,\mathcal{R} = \{\mathcal{R}_s \cup \mathcal{R}_e\},\mathcal{C})$ for each user $k$ where $n_s$ is the number of slots requested by the requesting UE. The set of candidate slots, $\mathcal{C} \subset \mathcal{S}$ is every slot within the allocation interval of the requesting UE with respect to a data message. The predictability of SPS transmissions will be utilized in the proposed schemes. A benefit of SPS transmission is that one resource allocation can be valid for multiple data message transmissions. Allocation of a SPS transmission is triggered at the \emph{trigger time}. The trigger time happens when the number of UEs within proximity is incremented to one (no longer zero) or after the resource re-selection counter expires. The resource re-selection counter is defined in \cite{3gpp_ts_2021} and decrements at each data transmission. The resource pool occupancy is given in the set $\{\mathcal{R}_s \cup \mathcal{R}_e\}=\{(s,o) | s \in \mathcal{S} ~\mathrm{and}~ o \in \mathbb{R}\}$ where $o_i$ is an indication of the occupancy defined as the strongest signal previously received from any of the UEs expected to transmit in slot $s_i$. If a slot $s_i$ is occupied by a UE within critical cooperation range of the requesting UE, the corresponding $o_i$ is set equal to infinity to avoid the half-duplex problem. $\mathcal{R}_e$ is provided by the allocating UE and $n_s$, $R_s$ and $S$ are provided by the requesting UE. $R_s$ indicates the current resource utilization as observed by the requesting UE whereas $\mathcal{R}_e$ indicates the resource utilization obtained (through control signaling) by the allocating UE.  If $K$ UEs are requesting an allocation from the same allocating UE simultaneously, their inputs will be ordered according to their priority, with $k=1$ indicating the highest priority UE and $k=K$ the lowest priority UE. 

Based on the resource occupancy from the requesting UE(s) and the received control signals, the resource allocation algorithm (Algorithm.~\ref{alg:RA}) allocates the resources for $\mathrm{UE}_k$ to avoid half-duplex problems and ensure the lowest interference from other UEs. If multiple requesting UEs are being assigned a resource allocation, the requesting UE with most potential half-duplex conflicts (most UEs in critical cooperation range) has resources allocated first. This greedy selection scheme is also know from greedy graph coloring algorithms. For each requesting UE, a set $\mathcal{P}$ of potential resources is initialized based on the resource pool occupancy observed by UE $k$. Resources are allocated based on the lowest occupancy in lines 4 and 5 of Algorithm~\ref{alg:RA}. In case multiple slots have identical minimum occupation in line 4, one will be randomly selected. A slot is allocated for UE $k$ in line 5 and the corresponding entry is removed from set $\mathcal{P}$ in line 6. As a result, the output of the resource allocation algorithm is the set of allocated resources,  $\mathcal{A}_k \subseteq \mathcal{C}_k$, for each of the requesting UEs.

\subsection{Device sequential resource allocation scheme}
\begin{figure*}[t]
	\centerline{\includegraphics[width=\linewidth]{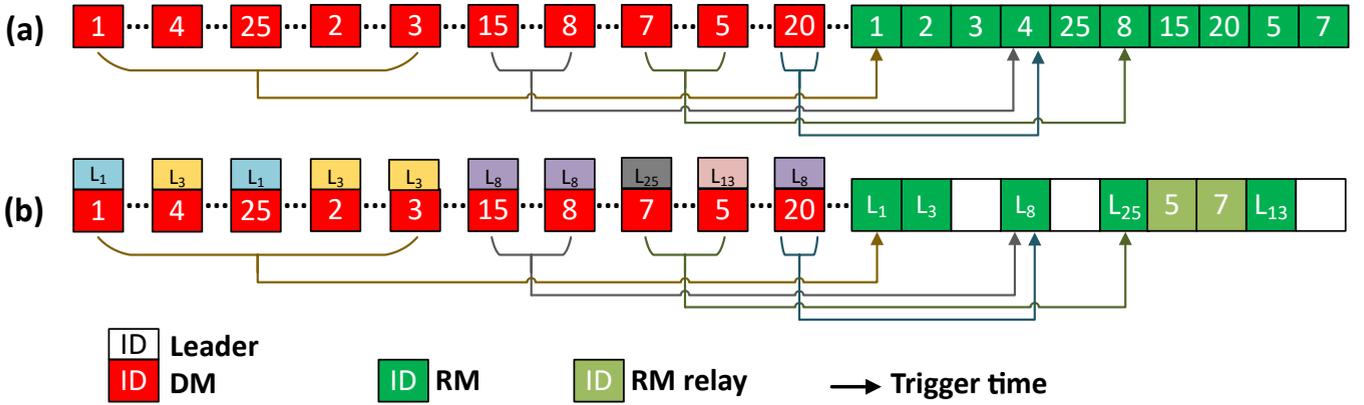}}
	\caption{Relation between discovery messages (DMs) illustrated by red boxes and resource selection messages (RMs) illustrated by green boxes for (a) device sequential and (b) group scheduling schemes. ID is UE identification.}
	\label{fig:signaling}
\end{figure*}
% rational behind the RA scheme and detailed coordination scheme
This scheme consists of coordinated resource selection by following a sequence in which UEs independently perform resource allocation in prioritized order. In our design, the UE priority is based on their trigger time and a unique ID. The UE with earliest trigger time has highest priority, and in case multiple UEs have identical trigger time, the unique UE ID determines the sequence such that lower ID has higher priority. 

\begin{figure}[t]
	\centerline{\includegraphics[height=8.1cm,keepaspectratio]{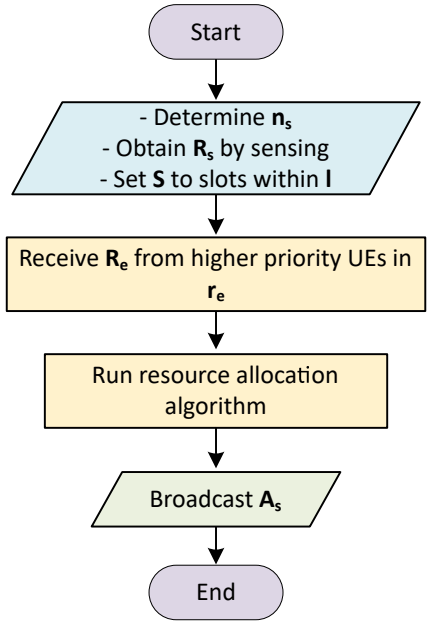}}
	\caption{Device sequential coordination scheme for $UE_i$}
	\label{fig:DevSeqFlowChart}
\end{figure}

In Fig. ~\ref{fig:signaling}~(a) the red boxes indicate the point in time when the trigger time is announced (in a discovery message discussed in Section \ref{sec:DM}). UEs 1, 4, 25, 2 and 3 have the same trigger time as indicated by the arrow pointing to the time slot for resource allocation. Due to the trigger time collision between the 5 UEs, they follow the prioritized order, which cause UEs 2, 3, 4 and 25 to perform resource allocation after their indicated trigger times (respectively a delay of 1, 2, 3 and 4 time slots).
% Coordination scheme
The coordination scheme for device sequential resource allocation follows the flow presented in Fig.~\ref{fig:DevSeqFlowChart}. A UE continuously monitor the trigger time and position of other UEs within the extended cooperation range $r_c$. Once the UE requires resources, it initiates the resource allocation scheme. After determining the number of resources necessary for the transmission, the UE-Awaits the resource selection from higher priority UEs within $r_c$ and continues when either resource selection has been received from all higher priority UEs or the \emph{resource selection delay} expires (further discussed in section \ref{sec:cntrlSig}). The resource selection delay is a configurable parameter. Then, the UE executes the resource allocation algorithm, providing itself with a resource allocation. The allocated resources is signaled by broadcast intended for every other UE within $r_c$. 

%Pro: autonomous, coordinated, prioritized
The advantage of the device sequential scheme is the autonomy with which each UE is performing its own resource allocation while simultaneously coordinating with UEs in extended coorperation range. Additionally, the prioritization scheme, while important for the coordination, is also a way of providing differentiated service to the swarm member UEs.
%Con: coordination delay, extra control signals
The potential drawback of the scheme is the additional resource allocation delay which might be incurred if the trigger time of multiple UEs within extended coordination range overlap, and the control overhead from the signals which indicate trigger time and selected resources between UEs. 

%This scheme consists of coordinated resource selection by following a sequential order based on the UE's unique ID. The lower the unique ID is, the higher the priority the UE has to select resources. Before resource allocation, UEs use both resource allocation of higher priority UEs and NR sidelink mode~2 sensing information. 
%The former includes the set of higher priority UE-Allocated resources obtained through message exchange within $r_c$. The latter contains the collection of convenient resources obtained in the procedure explained in Section \ref{sensingSL}. The combination of both allows UEs to avoid half-duplex problems among transmissions happening within $r_c$ or, in case of high grid occupancy, chose the one(s) that provide the lowest interference. The sequential procedure is presented in Fig.~\ref{fig:DevSeqFlowChart} where the resource allocation process is executed before obtaining the assigned resources. The detailed resource allocation procedure is shown in Fig.~\ref{fig:RA}.
%
%Sequential cooperation is not granted for free. Then, UEs must share their resource allocation with all higher priority ID UEs located within its $r_c$. Later, we will introduce a mechanism to distribute the necessary control signaling that enables sequential cooperation. 

\subsection{Group scheduling resource allocation scheme}
\begin{figure}[t]
	\centerline{\includegraphics[width=\textwidth,height=13.5cm,keepaspectratio]{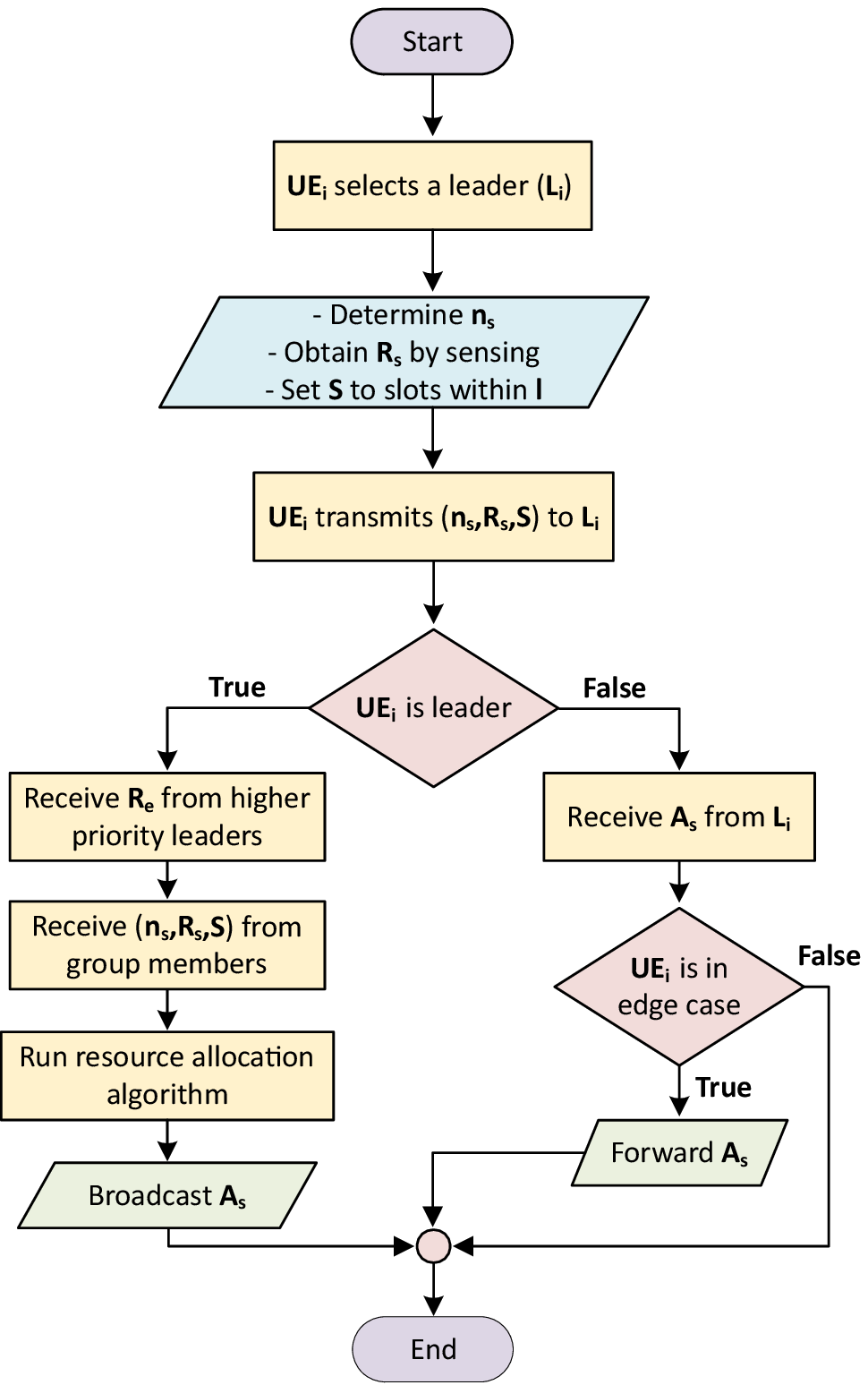}}
	\caption{Group scheduling coordination scheme for $UE_i$}
	\label{fig:GschFlowChart}
\end{figure}

%Rational
As implied by the naming, the group scheduling resource allocation scheme rely on local groups in which a group leader is executing the resource allocation algorithm and supplying the group members with resource allocation. Coordination happens within the group, but also between groups. For the latter, group leaders are either within extended cooperation range of, or have group members which collaborate with UEs in, another group. The group leader coordination is similar to the sequential scheme, where the prioritized order of allocation is determined by firstly group member trigger time (earlier trigger time is higher priority) and secondly the group leader unique ID (lower ID is higher priority). In Fig.~\ref{fig:signaling}~(b) UEs 1 and 3 have been elected as leaders with IDs $L_1$ and $L_3$, respectively. Both leaders have group members with the same trigger time (UEs 1 and 25 for $L_1$ and UEs 2, 3 and 4 for $L_3$). $L_1$ has highest priority, thus performs resource allocation for its group members before $L_3$. The flow of the group scheduling resource allocation scheme is presented in Fig.~\ref{fig:GschFlowChart}.

%Cooridnation scheme
A UE continuously maintain membership of a group. It does so by periodically performing leader selection and broadcasting the choice of group leader. The candidate leaders are all UEs within $r_e$. Out of the candidate leaders, the leader is chosen as the candidate with most UEs within $r_c$. The unique UE ID will resolve any such that the candidate leader with the lowest ID UE will selected as the leader. Thereby, the group leaders are bound to be involved in swarm communication.
Ahead of the trigger time, a UE informs its leader of its trigger time, the number of requested resources, the sensed resource occupation and the candidate slots.
At the trigger time the leader executes the resource allocation algorithm after receiving any potential resource selection from higher priority leaders within $r_e$, or at latest when the resource selection delay has expired. The output from the resource allocation is signaled to the requesting UE and any lower priority leaders within $r_e$.
Due to the range controlled leader selection procedure, the leaders of two collaborating UEs might be outside extended coordination range, potentially not being able to directly communicate. We refer to this as the \emph{edge case}. In edge cases, the collaborating UEs need to forward the resource allocation received from their leader to allow the leaders to coordinate the resource allocation. Such forwarding is performed by UE 5 and 7 in Fig.~\ref{fig:signaling}.

%Pro: speed, broader view, 
The advantage of the group scheduling scheme is that leaders are able to perform resource allocation for multiple UEs simultaneously and combining their allocation in a single control message, thereby reducing the amount of messages used for control signaling. Additionally, the group leader has more information for resource occupation as each group member and the group leader itself collects resource occupation information.
%Con
The disadvantages relate to the additionally required control signals. The resource allocation must be signaled between leader and requesting UE. Failure to receive this signal will cause the requesting UE to be without a resource allocation. The requesting UE must provide information to the leader which incurs additional signaling overhead. Lastly, the edge case where coordinating leaders are out of direct communication range will cause a coordination delay and additional overhead.

%This cooperative resource allocation scheme assembles on the conception of having group leaders who allocate resources for the whole group of UEs. Before resource allocation, UEs need to perform a leader selection procedure consisting of selecting the UE who has the highest number of UEs within its $r_c$. As this process is device-centric, a UE can be a group member, leader, or both. A particular case is presented when UEs have different group leaders but they are located within their respective $r_c$. Here, UEs share their particular leader selection such that unaware leaders will be conscious of other leaders' existence. It helps to avoid half-duplex problems caused by different leaders allocating the same resources to UEs located in the same $r_c$. The group scheduling resource allocation procedure is detailed in the flowchart shown in Fig.~\ref{fig:GschFlowChart}. When group leaders allocate resources for their group members, they follow the process previously shown in Fig.~\ref{fig:RA}.  

%In this case, cooperation requires sharing both leader selection (from UEs) and resource allocation (coming from group leader UE). Then, additional control signaling exchange is necessary. Next, a mechanism to do it will be proposed together with device sequential scheme. 
% Coordination: DSeq
% Coordination: GSch

\section{Control signaling for cooperative schemes} \label{sec:cntrlSig}
% By now it should be clear which information needs to be exchanged
The decentralized cooperative resource allocation schemes require additional control signaling exchanges compared to mode~2. In this section we establish the control messages which will carry the control signals. For the cooperative schemes we utilize all three methods in Fig.~\ref{fig:sigCategories} (RA control signaling embedded with data, embedded with independent control signaling and dedicated) for exchanging control signals. A summary of the control signals and their control information for each resource allocation scheme is presented in Table \ref{tab:msgs}. The data message is identical for all schemes and simply include an indication of the periodicity of the message, making any receiver able to determine future resource reservation. The next subsections will elaborate on the discovery and resource selection message types. 

\begin{table*}[]
	\caption{Message content necessary for the three resource allocation schemes}
	\label{tab:msgs}
	\resizebox{\textwidth}{!}{%
		\begin{tabular}{llll}
			& \cellcolor[HTML]{EFEFEF}\textbf{Mode~2} & \cellcolor[HTML]{EFEFEF} \textbf{Device sequential} & \cellcolor[HTML]{EFEFEF}\textbf{Group scheduling} \\
			\multicolumn{1}{l}{\cellcolor[HTML]{EFEFEF}\begin{tabular}[c]{@{}l@{}}\textbf{Discovery message} \\ \textbf{(DM) content}\end{tabular}}  & \cellcolor[HTML]{FFFFFF}\begin{tabular}[c]{@{}l@{}}\tabitem UE ID\\ \tabitem Position \& heading\end{tabular} & \cellcolor[HTML]{FFFFFF}\begin{tabular}[c]{@{}l@{}}\tabitem UE ID\\ \tabitem Position \& heading\\ \tabitem Trigger time\end{tabular} & \cellcolor[HTML]{FFFFFF}\begin{tabular}[c]{@{}l@{}}\tabitem UE ID\\ \tabitem Position \& heading\\ \tabitem Trigger time(s)\\ \tabitem Leader selection\\ \tabitem Sensed resource occupation $(R_s)$\\ \tabitem Special forward indication\end{tabular} \\

			\multicolumn{1}{l}{\cellcolor[HTML]{EFEFEF}\begin{tabular}[c]{@{}l@{}}\textbf{Resource selection} \\ \textbf{message (RM) content}\end{tabular}} & \multicolumn{1}{c}{\cellcolor[HTML]{FFFFFF}-} & {\cellcolor[HTML]{FFFFFF}\tabitem Resources allocated by UE $(A)$ } & {\cellcolor[HTML]{FFFFFF}\tabitem Resources allocated to group members $(A)$} \\

			\multicolumn{1}{l}{\cellcolor[HTML]{EFEFEF}\textbf{Data message}}  & \cellcolor[HTML]{FFFFFF}\begin{tabular}[c]{@{}l@{}}\tabitem Message periodicity\\ \tabitem Application data\end{tabular}  & \begin{tabular}[c]{@{}l@{}}\tabitem Message periodicity\\ \tabitem Application data\end{tabular} &    \cellcolor[HTML]{FFFFFF}\begin{tabular}[c]{@{}l@{}}\tabitem Message periodicity\\ \tabitem Application data\end{tabular}
		\end{tabular}%
	}
\end{table*}

\subsection{Discovery message (DM)} \label{sec:DM}

The objective of DMs is for UEs to become aware of each others ID, position and heading direction. It is transmitted periodically with no exceptions. The DM is necessary regardless whether the resource allocation scheme is cooperative or non-cooperative, e.g. mode~2. Each DM is scheduled randomly within the discovery period. %The mean period of DMs will be equal to the discovery period while the probability that a pair of UEs does not discover each other due to a half-duplex problem, $P(failed ~ DM)$, is

%\begin{equation}\label{eq:DMcol}
%	P( failed ~ DM )  = 1-\left(1 - \left( \frac{1}{n_d} \right) ^{m}\right)^{g}
%\end{equation}
%where $n_d$ is the number of slots within the discovery period, $m$ is the number of consecutive DM transmissions and $g$ is the number of devices within $r_e$.
%\begin{enumerate}
%	\item $n$ is the number of slots within the discovery period.
%	\item $m$ is the number of consecutive DM transmissions
%	\item $k$ is the number of devices within $r_e$
%\end{enumerate} 

%The objective of DMs is for UEs to become aware of each other position, and therefore each UE transmits it periodically. The DM is necessary regardless of the resource allocation scheme is cooperative or non-cooperative. Each DM is scheduled randomly within each discovery period. The mean period of DMs will be equal to the discovery period while the probability that a pair of UEs does not discover each other due to half-duplex problem is of a UE selecting overlapping resources () for $m$ consecutive DM transmissions with any of the $k$ other devices, $P(failed ~ DM)$, is
%P(collision ~ in ~ m ~ slots) = \left( \frac{1}{n} \right) ^{m}
%P(not colliding in m slots ) =  1 - \left( \frac{1}{n} \right) ^{m}
%P(not collidingin in m slots with any of k devices) = (1 - \left( \frac{1}{n} \right) ^{m})^k

For the device sequential and the group scheduling schemes the DMs are extended with information about the \textbf{trigger time}, when this is known by the UE. The UE can determine the trigger time either by estimating when another UE will be within $r_c$ or when the re-selection counter reaches zero. We assume that the trigger time can be estimated far in advance and that the minimum value of the reselection counter is 750 ms as specified in \cite{3gpp_ts_reselection}. A 100 ms discovery period would lead to each UE having at least 7 DM transmissions which results in sufficient discovery probability.
In the group scheduling scheme, the DM is extended with additional information. The leader selection is included in each discovery message such that leaders and collaborating UEs remain updated about the existing groups. When the trigger time approaches, the requesting UE will include the sensing result in its DM for the leader to use during resource allocation. Additionally, if UE-A identifies that its leader, $L_A$, and the leader, $L_B$, of a collaborating UE-B are out of direct communication range, UE-A will indicate in the DM the ID of $L_B$ and the trigger time of UE-B. This allows $L_A$ to determine the priority between the leaders and follow the coordination procedure. 
%DMs have a direct impact within two specific procedures: \textbf{leader selection} and \textbf{edge cases}. For the former, in addition to the trigger time, DMs serve an additional purpose by indicating the leader selection made by each UE. With this, the selected leader knows that it must allocate resources for the group member at the trigger time indicated in the DM. The leader includes this information within its DMs to enable cooperation with nearby leaders. Group members include the result of their sensing procedure in a DM, before their trigger time, to enable leaders to make the best allocation decision.
The size of DMs are enlarged by up to tenths of bytes due to the extensions needed by the cooperative schemes.

\begin{figure*}[t]
	\centerline{\includegraphics[width=\textwidth]{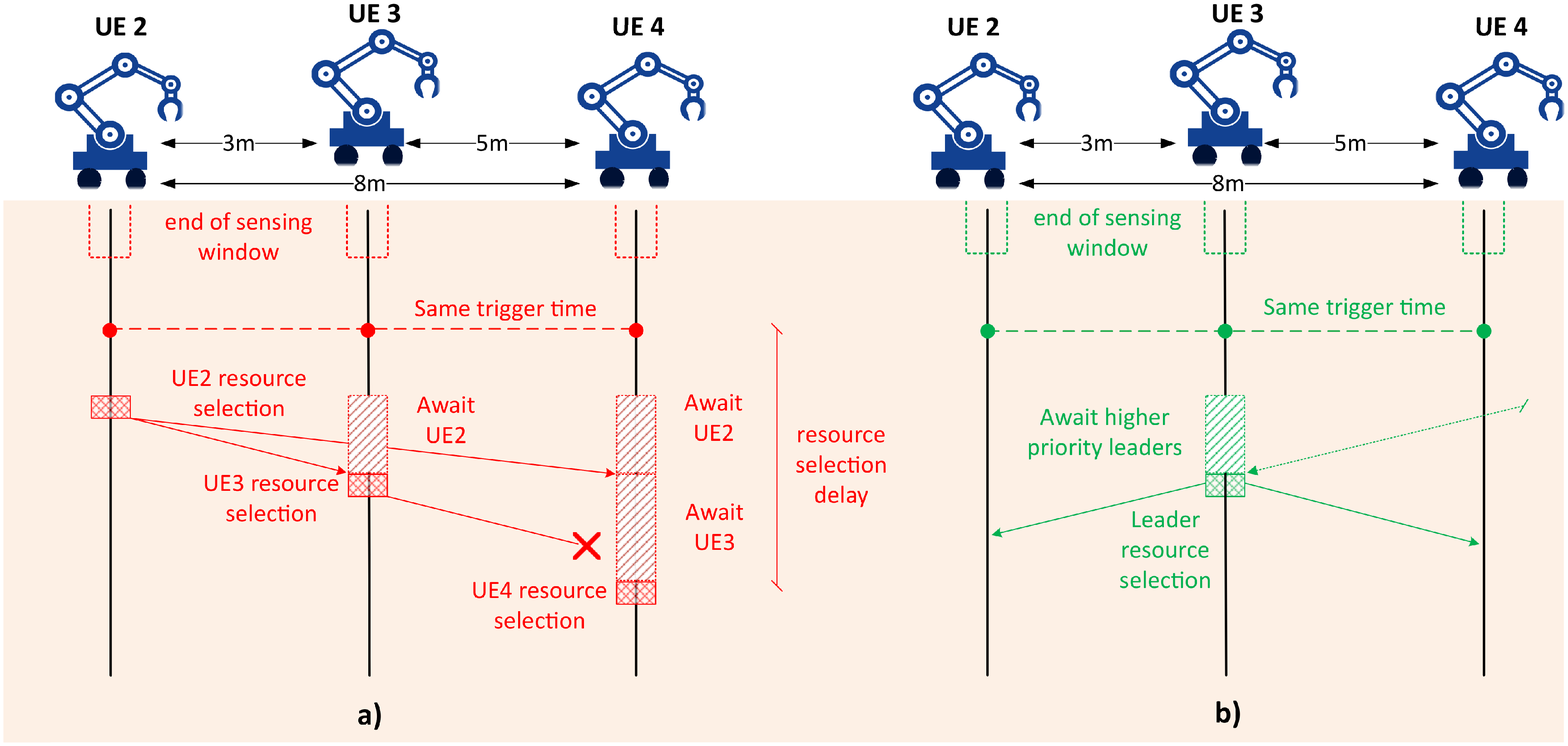}}
	\caption{Control signaling exchange for (a) device sequential and (b) group scheduling}
	\label{fig:CoordPhase}
\end{figure*}

\subsection{Resource selection message (RM)} \label{RMsubsection}
This control signal is exclusive to the cooperative resource allocation schemes. Its function is to carry information about the allocated resources for future data transmissions. Hence there is a direct connection between RM transmission and the trigger time indicated in the DMs. Compared to the non-cooperative scheme, the RM represents an additional overhead. However, it is transmitted only once per SPS period. Fig.~\ref{fig:CoordPhase} presents the RM transmission sequence diagram for each of the cooperative resource allocation schemes. The particularities of RMs for device sequential and group scheduling schemes are the following.

\subsubsection{Device sequential} \label{sec:devSeq}
At the trigger time a UE allocates resources and transmit its RM unless one of the two following conditions are true:
\begin{enumerate}
	\item there are UEs with higher priority (lower unique ID) within $r_e$ with the same trigger time (e.g. UE 3 waits for UE 2's RM in Fig.~\ref{fig:CoordPhase} (a)), or
	\item there are UEs within $r_e$, with earlier trigger time, which are pending to perform resource allocation (e.g., UE~8 is waiting for UE~25's RM in Fig.~\ref{fig:signaling} (a)).
\end{enumerate}
Therefore, upon reception of RMs from higher priority UEs or when the predefined resource selection delay has expired, the UE will proceed to send its RM. Even though the delay to perform resource allocation scales linearly with the number of higher priority UEs in the sequence, it is bound by the resource selection delay. Resource allocation commences once the resource selection delay expires (e.g., resource selection delay expires for UE~4 and it performs its resource allocation as illustrated in Fig.~\ref{fig:CoordPhase} (a)).

\subsubsection{Group scheduling}
%In case $L_B$ has higher priority than $L_A$, UE-A will forward the resource selection message from $L_B$ to $L_A$ if it is received within the resource selection delay.
RMs are transmitted from the group leaders to their respective group members at the trigger time. If two or more leaders within $r_e$ have inferiors with the same trigger time (e.g. the leader, UE~3, waits for higher priority leaders resource allocation in Fig.~\ref{fig:CoordPhase} (b)), they must follow the sequential procedure explained in Section \ref{sec:devSeq}. In cases where multiple group members have been given resources simultaneously (e.g. UEs 2, 3, and 4 in Fig.~\ref{fig:CoordPhase} (b)), the group leader combines the selected resources in one RM. This is beneficial in dense scenarios since it reduces the load of control signals in the control resource pool in comparison to the device sequential scheme.
For the special forwarding procedure (edge case), group member UEs should forward the RMs (e.g. UEs 5 and 7 in Fig.~\ref{fig:signaling} (b)) between leaders to enable leader-cooperation and hence, avoid half-duplex problems when allocating resources within their respective groups. 
The delay to perform resource allocation in the group scheduling scheme scales with the number of leaders within $r_e$ of each other. In addition, the special forwarding procedure introduces the delay of up to two additional transmission times. However, initiation of the resource allocation is bounded by the configurable resource selection delay.% Overhead: time addition scales with leader density plus edge cases

\section{System Level Evaluation} \label{sec:sim}
%Swarm communication has potential to gain a footage across different industries. For at concrete evaluation scenarios, we focus on the manufacturing industry which is facing a transition to the fourth industrial era (industry 4.0) characterized by the adaption of "smart" technology \cite{kumar_industry_2019}, where large scale communication is fostering new ways to utilize data and improved production flexibility. Production flexibility is necessary to maintain a cost and resource efficient production \cite{kruger_manufacture_2017}, as costumer demand becomes increasingly customized. The flexible production contemplates breaking the current linear (conveyor belt) production concept into a nonlinear one in the form of swarm production where products are transported by autonomous mobile robots (AMRs) between the different stations across the factory \cite{rodriguez_5g_2021}. The swarm of AMRs must collaborate to achieve peak production output and mitigate collisions.

We consider an application for collective environment perception, in which robots within a proximity of $r_c = 5$ m must establish real-time high-throughput communication at high reliability for cooperative behavior, e.g. collision avoidance among robots and with external objects. This scenario is not unlike collective perception and cooperative collision avoidance use cases from vehicle to anything (V2X) envisioned by 3GPP in \cite{3gpp_tr_2018}. Specific requirements for this scenario are a $10$ Mbps throughput where message latency does not exceed $10$ ms at a reliability of $99.99$ \% \cite{3gpp_tr_2018}.

% Mobility and channel...
The robots are driving in a rectangular indoor factory building. Each robot moves according to the random waypoint mobility model in which the robot moves at fixed speed between random points within the factory. The 3GPP non-line of sigth indoor factory with sparse clutter and low base station (InF-SL) pathloss model from \cite{pathloss_3gpp} is used for modeling the pathloss on links. UE antennas are omnidirectional. As multiple links are in use, we impose correlation on the shadowing component. The shadowing is computed according to the method in \cite{efficient_correlated_shadowing_modelling} where integration over a Gaussian random field enforces a $20$ m de-correlation distance and $5.7$ dB standard deviation. Fast fading is not explicitly modeled, but included in the link layer model.

% Numerology, bandwidth, interference
Regarding 5G NR parameters we select numerology 2, dictating a $d_s= 0.25$ ms slot duration. The data channel bandwidth is $100$ MHz whereas the control data is carried on the smallest configurable sidelink sub-channel of twelve sub-carriers resulting in a $7.2$ MHz bandwidth. The lowest modulation and coding scheme (MCS) for sidelink has modulation order $2$ and coderate $\frac{120}{1024}$, leaving at most 196 bits for the control messages. %(the smallest sub-channel size for sidelink transmission is 10 resource blocks each consisting of 12 sub-carriers) % Each control message wil be able to carry (at the lowest MCS of mod-order 2 and coderate 120/1024) 196 bits per slot (10 resource blocks * 12 sub-carriers pr RB * 14 OFDM symbols = 1680 resource elements. Each resource element is modulated with one bit, and due to the coderate 1680*120/1024 = 196 bits in the slot) 
The MCS for the data transmission is dynamically adapted at the time of allocation.  For each robot within $r_c$, the signal to interference and noise ratio (SINR) is measured on the most recent transmission. The worst SINR is used to determine the modulation and coding scheme from \cite[Table~5.1.3.1-2]{RA_3gpp} which can attain a 0.01~\% target block error rate (BLER). The link level, hence the mapping from SINR to BLER, is modelled using a set of BLER curves generated from separate link level simulations \cite{lagen_new_2020}. The link level simulation includes all physical layer processing according to 5G NR. The required number of slots $n_s$ are calculated based on the selected MCS, assuming that the transport block is bit padded to an integer number of slots. We do not differentiate between data and control signal transmission in the link level modeling which makes the control link performance slightly optimistic due to the much lower transmission bandwidth, i.e. 100MHz compared to 7.2MHz. Simulations parameters are listed in Table \ref{tab:params}.

\subsection{Key performance indicators}
% Explain the half duplex issues and the types of interference which might incur. Also talk about reasons why errors might occur due to missing transmission (rm-retransmission, transmission suppresion/deprivation)
The main key performance indicator is reliability - the probability that a data message is received within the latency constraint. We measure it in the form of failure probability. The target 99.99\% reliability corresponds to a $10^{-4}$ failure probability. As a complementary key performance indicator we capture the packet inter-reception (PIR) metric defined by 3GPP in \cite{3gpp_tr_PIR_2019}. It indicates the time in between successive packet receptions and is important for applications where regular updates are required. 
Multiple reasons might cause a reception failure, e.g. half-duplex errors arise when a UE is transmitting and therefore not able to receive a data transmission. We differentiate between whether a UE is transmitting a data message (half-duplex data), a discovery message (half-duplex DM), or a resource selection message (half-duplex RM). 

%The purpose of all the evaluated resource allocation schemes is to avoid half duplex issues caused by transmission of data. Half-duplex caused by DM can be reduced based on the non-overlapping technique. RMs are much less frequent than DM and data transmissions and immediate transmission of RMs is necessary to meet the data transmission latency.

Interference is another source of data reception failure. We differentiate between interference caused by UEs within $r_c$, denoted \emph{inner} interference, and interference by UEs outside $r_c$, denoted \emph{outer} interference. When UEs within and outside $r_c$ simultaneously cause interference we denote it as \emph{mixed} interference.

Lastly, when a group member has not received the resource selection message from its leader (no RM reception), it cannot perform a data transmission which will cause data reception failures at the receivers.% or the resource allocation was suppressed/deprived.
%We count an error for each intended receiver which did not receive the expected transmission.

%\subsection{Simulation configurations}
%
%For the evaluations, five configurations were performed. 
%\begin{enumerate}
%	\item \textbf{Error-free signaling} in which every control message is received at every intended receiver.
%	\item \textbf{Error-prone signaling} according to the prevailing signal conditions.
%	\item \textbf{Signaling plus the RM re-transmission technique} in which in addition to 2) the RM re-transmission technique is utilized in the group scheduling scheme to mitigate data failures caused by failure to receive RMs.
%	\item \textbf{Signaling plus the RM re-transmission and the non-overlapping technique} in which in addition to 3) the non-overlapping technique is utilized to schedule DM in time slots where no incoming data transmissions are expected.
%	\item \textbf{Signaling plus the non-overlapping and piggybacking techniques} in which in addition to 4), the piggybacking mechanisms are enabled for the cooperative RA schemes.
%\end{enumerate}
%
%The mobility trace of each swarm size was reused for all five configuration to allow direct comparison between the configurations. A 1000 seconds simulation time allows the robots to traverse the facility multiple times causing various collaborative configurations. Simulations parameters are listed in Table \ref{tab:params}.

\begin{table}[t]
	\caption{Simulation parameters}
	\label{tab:params}
	\begin{tabularx}{\linewidth}{l L}
		\hline
		Parameter                                  & Value/range                                                                               \\ \hline
		Carrier frequency, $f_{\textrm{c}}$        & $3.5$ GHz                                                                                 \\
		Swarm size (number of UEs)                 & $[10,20,30,40,50,60,70]$                                                                     \\
		Critical cooperation range, $r_{\textrm{c}}$& $5$ m                                                                                    \\
		Extended Cooperation range, $r_{\textrm{e}}$& $25$ m                                                                                   \\
		Facility dimensions                        & $120 \times 50 ~ {\textrm{m}^2}$    \cite{pathloss_3gpp}                                  \\
		Transmission power, $P_{\textrm{tx}}$      & $0$ dBm                                                                                   \\
		Data channel bandwidth                     & $100$ MHz                                                                                 \\
		Control channel bandwidth				   & $7.2$ MHz																				   \\
		Resource selection delay                   & $1.25$ ms                                                                                 \\
		NR slot duration						   & $250$ $\mu$s																			   \\
		Thermal noise power spectral density       & $-174$ dBm/Hz             	                                                               \\
		Receiver noise figure				       & $9$ dB                      	                                                           \\
		Interference                               & Independent intra-system 				                                                   \\
		UE speed                                   & $1$ m/s                                                                                   \\
		Mobility model							   & Random waypoint (RWP) 		            										           \\
		Pathloss model							   & InF-SL	\cite{pathloss_3gpp}									           				   \\
		Propagation condition					   & Non line of sight																		   \\
		De-correlation distance $\delta$           & $20$ m \cite{efficient_correlated_shadowing_modelling}                                    \\
		Discovery message periodicity              & $100$ ms                                                                                  \\
		Data message periodicity, $d_p$                   & $10$ ms                                                                                   \\
		Data message size, $x_d$                        & $100$ kb                                                                                  \\
		Data message latency requirement, $l$                        & $10$ ms                                                                                  \\
		Simulation time                            & $1000$ s                                                                                  \\ \hline
	\end{tabularx}%
\end{table}

\begin{figure}[t]
	\centerline{\includegraphics[width=0.5\textwidth]{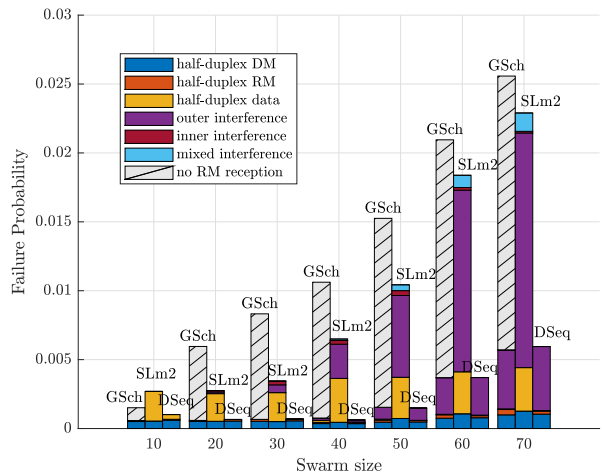}}
	\caption{Failure probability and the causes of data transmission failures (half-duplex of DM, RM and data, inner, outer and mixed interference, and no RM reception) for three resource allocation schemes (SLm2, DSeq and GSch)}
	\label{fig:causes}
\end{figure}

\section{Simulation Results} \label{sec:eval}

The control signaling exchange has a direct impact on the data exchange performance. It is fundamental to fulfill two conditions. First, the random selection of DM transmission must not coincide with the reception of data transmissions since it will cause half-duplex problems (hald-duplex DMs). Second, RMs failure probability should be sufficiently low such that it does not inhibit the performance of the cooperative schemes.  

\subsection{Reliability analysis and enhancement techniques}
In Fig.~\ref{fig:causes} we present the failure probability and the causes at various swarm sizes for the three resource allocation schemes group scheduling (GSch), mode~2 (SLm2) and device sequential (DSeq).

\begin{figure}[t]
	\centerline{\includegraphics[width=0.5\textwidth]{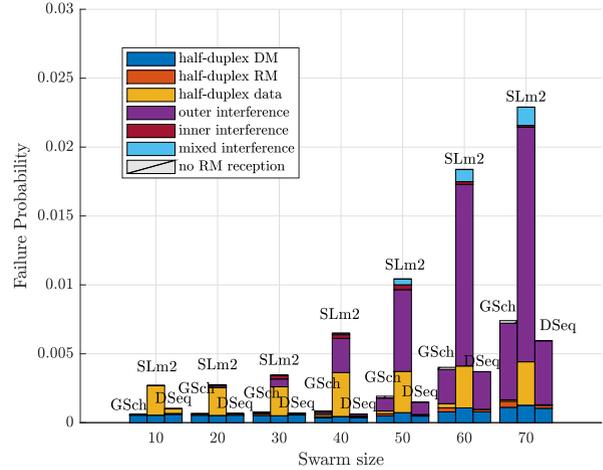}}
	\caption{Failure probability and the causes of data transmission failures for three resource allocation schemes after enabling RM re-transmissions}
	\label{fig:RMreTx}
\end{figure}

%Justifying the RM-retransmission 
Failure to receive RMs in the cooperative resource allocation schemes can result in non-cooperative resource allocation. In the group scheduling scheme, a group member UE is dependent on receiving the RM from its leader. Failure of this RM reception will result in failure to transmit data for the entire SPS data transmission period (grey hatched bars in Fig.~\ref{fig:causes}). To address this problem the \emph{RM re-transmission} technique was incorporated. It enables the group member to send a non-acknowledgment (NACK) to its leader indicating that re-transmission of the RM is necessary. It might take several NACKs for successful reception of RM. Fig.~\ref{fig:RMreTx} illustrates how failures caused by no RM receptions diminishes.

\begin{figure}[t]
	\centerline{\includegraphics[width=0.5\textwidth]{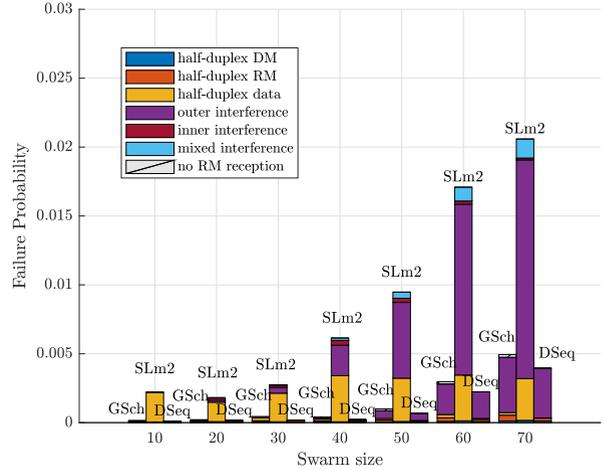}}
	\caption{Failure probability and the causes of data transmission failures for three resource allocation schemes after enabling RM re-transmissions and non-overlapping techniques}
	\label{fig:nonOverlapping}
\end{figure}

%Justifying the non-overlapping technique
The second largest failure cause (at small swarm sizes) is half-duplex failures caused by transmission of discovery messages (blue bars in Fig.~\ref{fig:causes}). The random transmission of DMs has a significant impact on total failure probability of the cooperative resource allocation schemes. Mode 2 is similarly affected by the half-duplex DM. To counteract this problem we propose the \emph{non-overlapping} technique. It utilizes the information about the current SPS transmissions acquired by UEs during the sensing procedure. The SPS transmission slots acquired by other UEs are not considered as possible options for the transmission of DMs to reduce potential half-duplex problems. Fig.~\ref{fig:nonOverlapping} depicts the near disappearance of half-duplex DM failures. 

% Piggy
Additionally, a few half-duplex problems occur in receiving data due to simultaneous data transmission (yellow bars for Gsch in Fig.~\ref{fig:causes}) even at small sizess. This indicates that leaders were not cooperating. In the 40 UE swarm size, the device sequential scheme also experiences half-duplex failures to receive data due to simultaneous data transmissions (half-duplex data). This is an indication that some UEs failed to follow the sequential procedure. These described issues lead to the application of the \emph{piggybacking} technique for the respective resource allocation schemes. Piggybacking builds on repeating the resource selection information by appending it to other RMs. It is done as follows in the two cooperative schemes:

\begin{itemize}
	\item [$-$] \textit{Device sequential:} When a UE receives RMs from its predecessors, it includes this information in its respective RM, so that if UEs that follow the sequence did not receive previous RMs, they can recover them. 
	\item [$-$] \textit{Group scheduling:} When the group leader sends an RM to a group member UE, it includes the information of prior transmitted RMs. It allows group member UEs an additional chance to receive its resource allocation when the leader schedules other inferiors
\end{itemize}

\begin{figure}[t]
	\centerline{\includegraphics[width=0.5\textwidth]{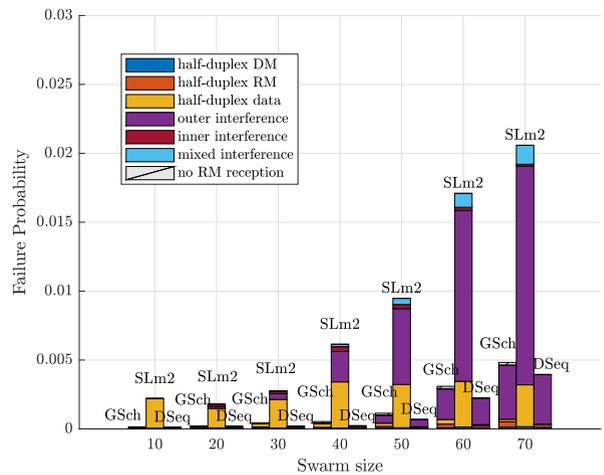}}
	\caption{Failure probability and the causes of data transmission failures for three resource allocation schemes after enabling RM re-transmissions, non-overlapping and piggybacking techniques}
	\label{fig:fullFeatures}
\end{figure}

Fig.~\ref{fig:fullFeatures} illustrates that the effect of the piggybacking is negligible. This is a sign that the allocation sequences are not long enough even at swarm sizes of 70 UEs for the piggybacking technique to have an effect.

At large swarm sizes, outer interference becomes the main cause of failure. We plan to address it in our future work.

%Difference between mode~2 and the coordinated RA schemes
Even without the improvement techniques enabled, the device sequential RA scheme outperforms mode~2. At small swarm sizes the main difference lies in mode~2 having a considerable number of half-duplex data failures. The half-duplex failures caused by transmission of RMs in the cooperative schemes constitute a minor performance impact. As swarm size increases, interference becomes a dominant failure cause. The lack of cooperation and the resource selection procedure of mode~2 (described in \ref{sec:resourceSelection}) result in UEs experiencing data reception failures caused by high interference coming from UEs outside cooperation range (outer interference), UEs inside cooperation range (inner interference), or both (mixed interference).

%We note that the inferiors failing to receive RMs from their leader (no RM reception) are detrimental to the performance of the group scheduler. These failures represent the main reason that the group scheduling scheme has the highest failure probability of the three RA schemes. By applying the RM re-transmission technique these failures are largely mitigated for the group scheduling scheme. 

\subsection{Reliability performance with enhancements} \label{sec:failureProb}
\begin{figure}[t]
	\centerline{\includegraphics[width=0.5\textwidth]{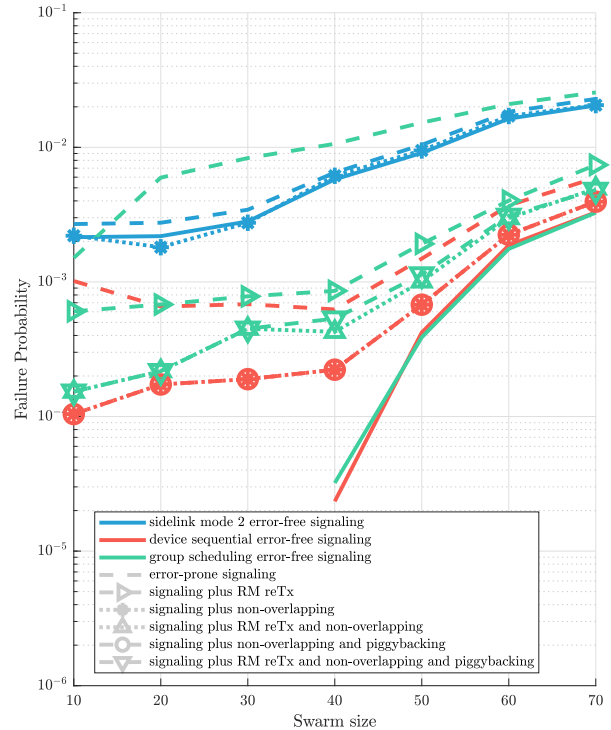}}
	\caption{Failure probability for the resource allocation schemes at the five simulation configurations: error-free signaling, error-prone signaling, error-prone signaling with re-transmissions (only for group scheduling scheme), error prone with non-overlapping (re-transmissions), and error-prone signaling with non-overlapping plus piggybacking (re-transmissions)}
	\label{fig:errorProb}
\end{figure}

%-Plot 2
%	-Impact of signaling to the Perfect signaling approach
%	-Refer to the failures causes in previous plot, so we see that the applied techniques mitigate those problems and reliability is increased
%-Compare RA schemes
%	-device sequential is better than group scheduling (mode~2 in the middle)
%	-How to improve group scheduling
%	-tendencies of RA schemes (group tends to reduce failure probability as load increases)
Fig.~\ref{fig:errorProb} shows the failure probability for different swarm sizes in simulations with the following configurations:
\begin{enumerate}
	\item \textbf{Error-free signaling} in which every control message is received at every intended receiver.
	\item \textbf{Error-prone signaling} according to the prevailing signal conditions.
	\item \textbf{Signaling plus the RM re-transmission technique} in which in addition to 2) the RM re-transmission technique is utilized in the group scheduling scheme to mitigate data failures caused by failure to receive RMs.
	\item \textbf{Signaling plus the RM re-transmission and the non-overlapping technique} in which in addition to 3) the non-overlapping technique is utilized to schedule DMs in time slots where no incoming data transmissions are expected.
	\item \textbf{Signaling plus the non-overlapping and piggybacking techniques} in which in addition to 4), the piggybacking techniques are enabled for the cooperative RA schemes.
\end{enumerate}
Mode~2 (blue lines in Fig.~\ref{fig:errorProb}) reaches failure probability below $10^{-2}$ until swarm size of 50. The failure probability of mode~2 is barely affected by the simulation configuration. The highest failure probability is observed in the error prone signaling configuration, where in addition to half-duplex data and interference, errors were caused by half-duplex DM. Enabling the non-overlapping technique brings the error probability of mode~2 down to the level of error free control signaling.

Device sequential resource allocation is affected by the enhancement techniques. With error prone signaling the failure probability below $10^{-3}$ can be maintained until swarm size of 40. Enabling the non-overlapping technique further reduces the failure probability and allow it to maintain failure probability below $10^{-3}$ until 50 UE swarm sizes. The piggybacking technique has no impact. The device sequential scheme is able to meet the $10^{-4}$ failure probability target at 10 UE swarm size when all enhancement techniques are enabled. With error-free signaling, the device sequential scheme experiences no failures at swam sizes smaller than 40 UEs.  

Group scheduling with error prone signaling has the highest failure probability of all schemes and configurations due to the impact on non-received RMs. However, enabling RM re-transmissions reduces the failure probability by an order of magnitude and makes the performance comparable to the device sequential scheme in the error prone signaling configuration. Enabling non-overlapping further reduces the failure probability of the group scheduling scheme maintaining the failure probability below $10^{-3}$ untill swarm size of 50. With all features enabled the failure probability of group scheduling is still slightly higher than that of device sequential. With error free signaling, the group scheduling performance is as good as device sequential.

\subsection{Packet inter reception (PIR)}
\begin{figure*}[t]
	\centerline{\includegraphics[keepaspectratio,width=\textwidth]{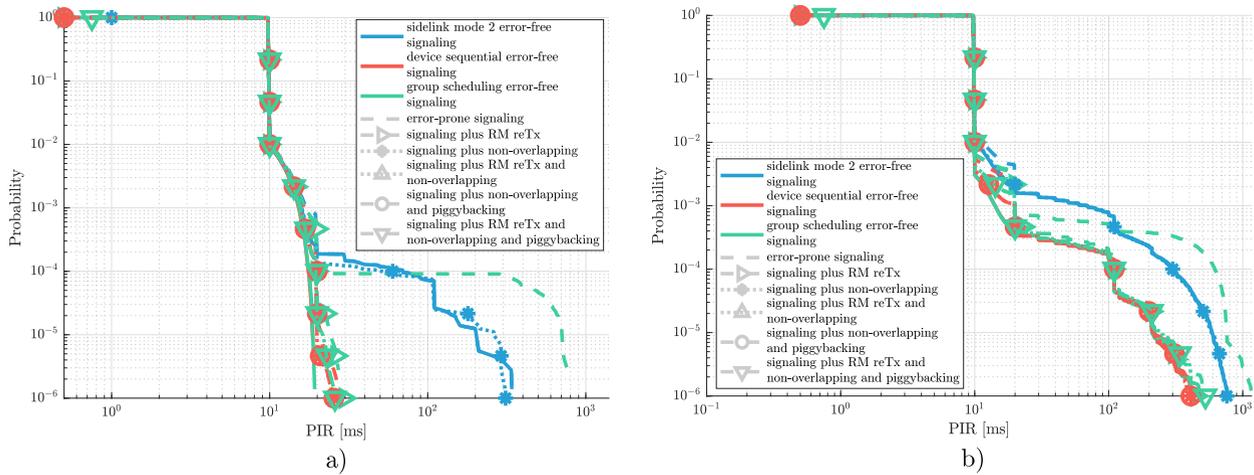}}
	\caption{Packet inter reception (PIR) at swarm sizes of 20 UEs (a) and 70 UEs (b) for all simulation configurations}
	\label{fig:pir}
\end{figure*}

Fig.~\ref{fig:pir} (a) and (b) show the complementary cdf of the PIR for respectively 20 and 70 UE swarm size simulations. At both loads a PIR less than or equal to 10 ms is most frequent. This is expected, as the SPS period is exactly 10 ms, thus successive successful receptions of data messages in the same series of SPS transmissions will result in a 10 ms PIR. A PIR lower than 10 ms can occur as a result of re-selection of SPS transmission, and the same goes for PIR between 10 and 20 ms. However, PIRs longer than 20 ms are caused by reception failures. The configuration with the highest failure probability also experience the longest PIR, regardless of allocation scheme. At 20 UE swarm size, the PIRs exceed 20 ms with a probability less than $10^{-3}$. Only mode 2 and the group scheduling configuration with error prone signaling experience 30 ms, corresponding to 2 successive reception failures. At 70 UE swarm size all configurations experience PIRs greater than hundreds of milliseconds. The cooperative schemes perform similar in configurations with RM-retransmissions enabled and outperform mode 2 at both swarm sizes.

%\begin{figure}[t]
%	\centerline{\includegraphics[width=0.5\textwidth]{Figures/PIR20UES}}
%	\caption{Packet inter reception (PIR) at swarm size of 20 UEs for all simulation configurations}
%	\label{fig:pir20}
%\end{figure}
%
%\begin{figure}[t]
%	\centerline{\includegraphics[width=0.5\textwidth]{Figures/PIR70UES}}
%	\caption{Packet inter reception (PIR) at swarm size of 70 UEs for all simulation configurations }
%	\label{fig:pir70}
%\end{figure}

% !TeX spellcheck = en_US
\section{Conclusion} \label{sec:con}

5G NR sidelink mode~2 is the current baseline resource allocation scheme for swarm communication. However, the autonomy of mode~2 and its random resource allocation algorithm is an impediment for its ability to accommodate the growing demand for high performance in dense swarms. We proposed two cooperative resource allocation schemes - \emph{device sequential} and \emph{group scheduling} - each representing a different coordination scheme.%s which facilitate a heuristic resource allocation algorithm.

We evaluate the proposed resource allocation schemes against baseline mode~2 in a series of comprehensive system level simulations. Despite the increased signaling overhead necessary in the coordinated schemes, they still represent an order of magnitude reduction in failure probability when compared to mode~2.

%Even though signaling is imposing multiple orders of magnitude reliability decrease on our proposed cooperative resource allocation schemes they maintain an order of magnitude improvement over the baseline sidelink mode~2. 
%- i.e. the tendency of the cooperative gain identified in paper 1 is maintained for device sequential despite signaling. Group scheduler suffers from unsuccessful receptions of the resource selection messages (from leader to inferior).

The methodology of identifying distinct causes of data failure provided valuable insight. Three enhancement techniques, respectively, resource selection message re-transmissions, non-overlapping allocation of discovery messages and piggybacking, were designed to address the data transmission failures caused by the error prone control signaling. Resource selection message re-transmission and non-overlapping allocation of discovery messages proved to significantly reduce failure probability in the coordinated schemes, whereas piggybacking did not introduce any significant gain. 

%\subsection{Future work}
The proposed resource allocation schemes, their associated control signaling and enhancement techniques provide a good trade-off between control overhead and performance in terms of latency and reliability. However, in order to achieve the stringent 99.99\% reliability requirement additional interference management techniques are necessary. In our future work we will explore techniques to improve the reliability at larger swarm sizes.

%Interference becomes the dominant failure cause when load increases, and therefore we plan to address these failures in our future work.

%The two proposed techniques serves to address the specific problems caused by the signaling. They work. Most significant is the non-overlapping mechanism (discovery not transmitted while data is being received)

%At low loads the cooperative resource allocation (device sequential) outperforms regular mode~2 by a factor of 10. Due to the HD-data issues of mode~2. I.e. cooperation is able to deal with the main problem of mode~2 which can lead to HD-data issues

%The normalized area spectral efficiency (NASE) shows that (hopefully) the cooperative schemes gets closer to the maximum achievable NASE than mode~2, i.e. more UEs can be served in the same area. Other (lower layer) techniques could build on top of these gains, e.g. spatial diversity (beamforming,MIMO,...)

%Reasons for failures have been highlighted: they will serve as guidance for future improvements (highlights specific problems to tackle) for the schemes. Aim is to improve the reliability to four nines 99.99\% in future studies.

\bibliographystyle{IEEEtran}
\bibliography{Bibliography2}

\begin{IEEEbiography}[{\includegraphics[width=1in,height=1.25in,clip,keepaspectratio]{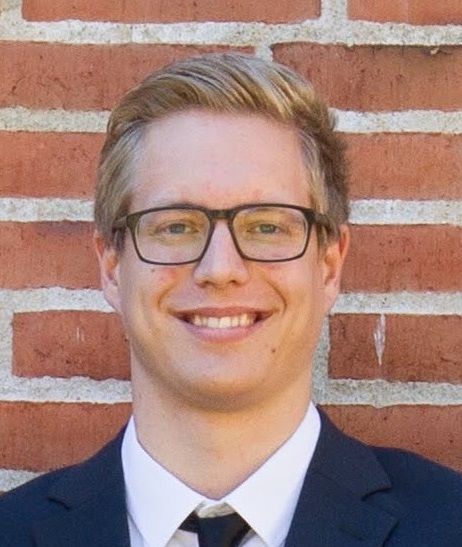}}]{Rasmus L. Bruun} received his B.Sc. in engineering (internet technologies and computer systems) and M.Sc. in engineering (networks and distributed systems) from Aalborg University, Denmark in 2015 and 2018 respectively. He is currently pursuing a Ph.D. degree in the Wireless Communication Networks section at Aalborg University in collaboration with Nokia Standardizations at the Aalborg, Denmark office. His research activities include mobile wireless ad-hoc networks, propagation modeling and radio resource management.
\end{IEEEbiography}
\begin{IEEEbiography}[{\includegraphics[width=1in,height=1.25in,clip,keepaspectratio]{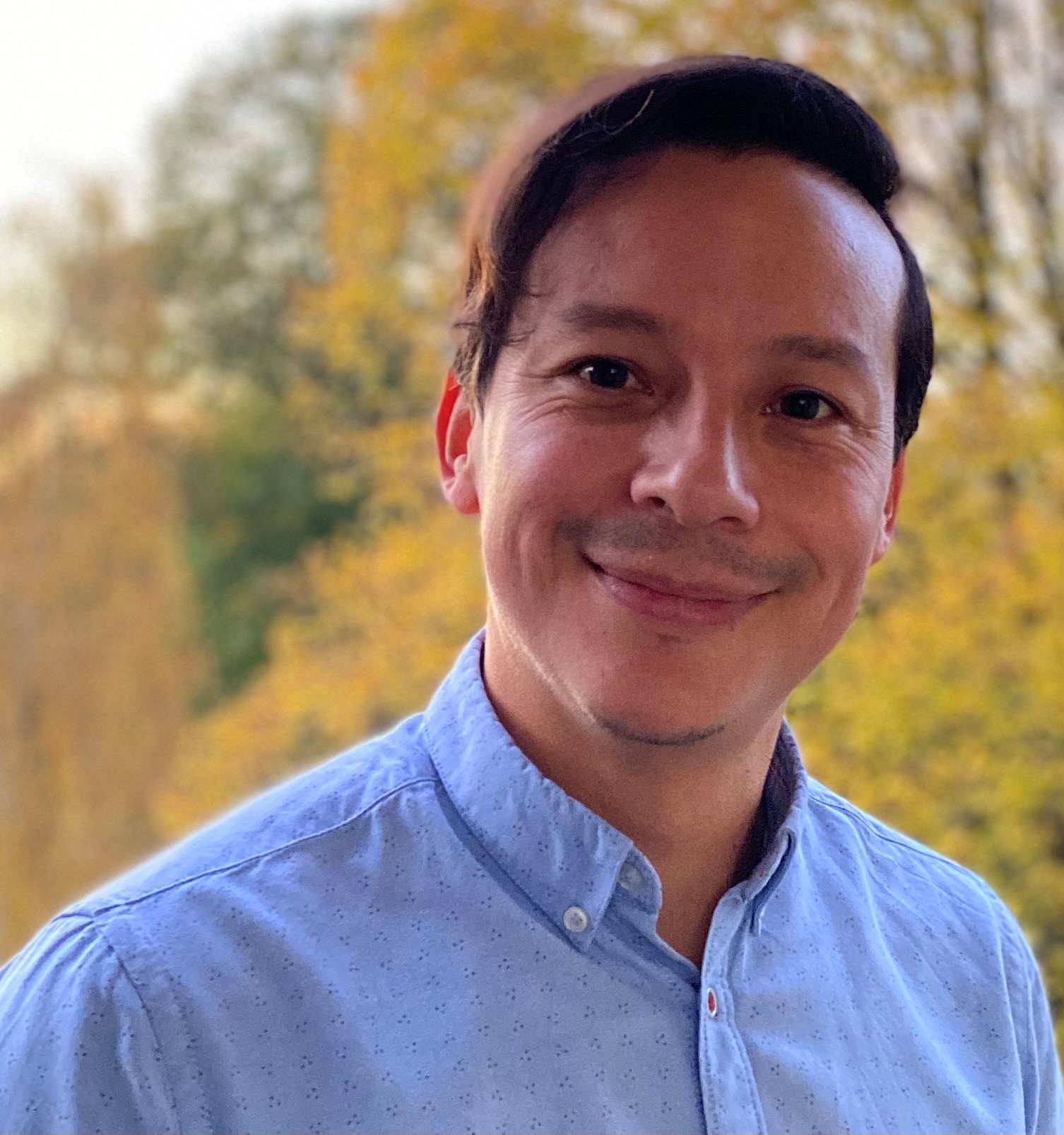}}]{C. Santiago Morejón García} received his Eng. degree in electronics and telecommunications from Escuela Politécnica Nacional, Ecuador in 2012 and his M.Sc. in mobile communications from Telecom ParisTech/Eurecom, France in 2017. He is currently pursuing a Ph.D. degree in the Wireless Communication Networks section at Aalborg University in collaboration with Nokia Standardizations at the Aalborg, Denmark office. His research activities include radio resource management and interference mitigation for decentralized networks using 3GPP sidelink. 
\end{IEEEbiography}
\begin{IEEEbiography}[{\includegraphics[width=1in,height=1.25in,clip,keepaspectratio]{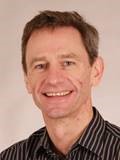}}]{Troels B. Sørensen} received the Ph.D. degree in wireless communications from Aalborg University in 2002. Upon completing his M.Sc. E.E. degree in 1990, he worked with type approval test methods as part of ETSI standardization activities. Since 1997 he has been at Aalborg University, where he is now Associate Professor in the section for Wireless Communication Networks. His research and teaching activities include cellular network performance and evolution, radio resource management, propagation characterisation and related experimental activities. He has successfully supervised more than 15 PhD students, and published more than 120 journal and conference papers. 
\end{IEEEbiography}
\begin{IEEEbiography}[{\includegraphics[width=1in,height=1.25in,clip,keepaspectratio]{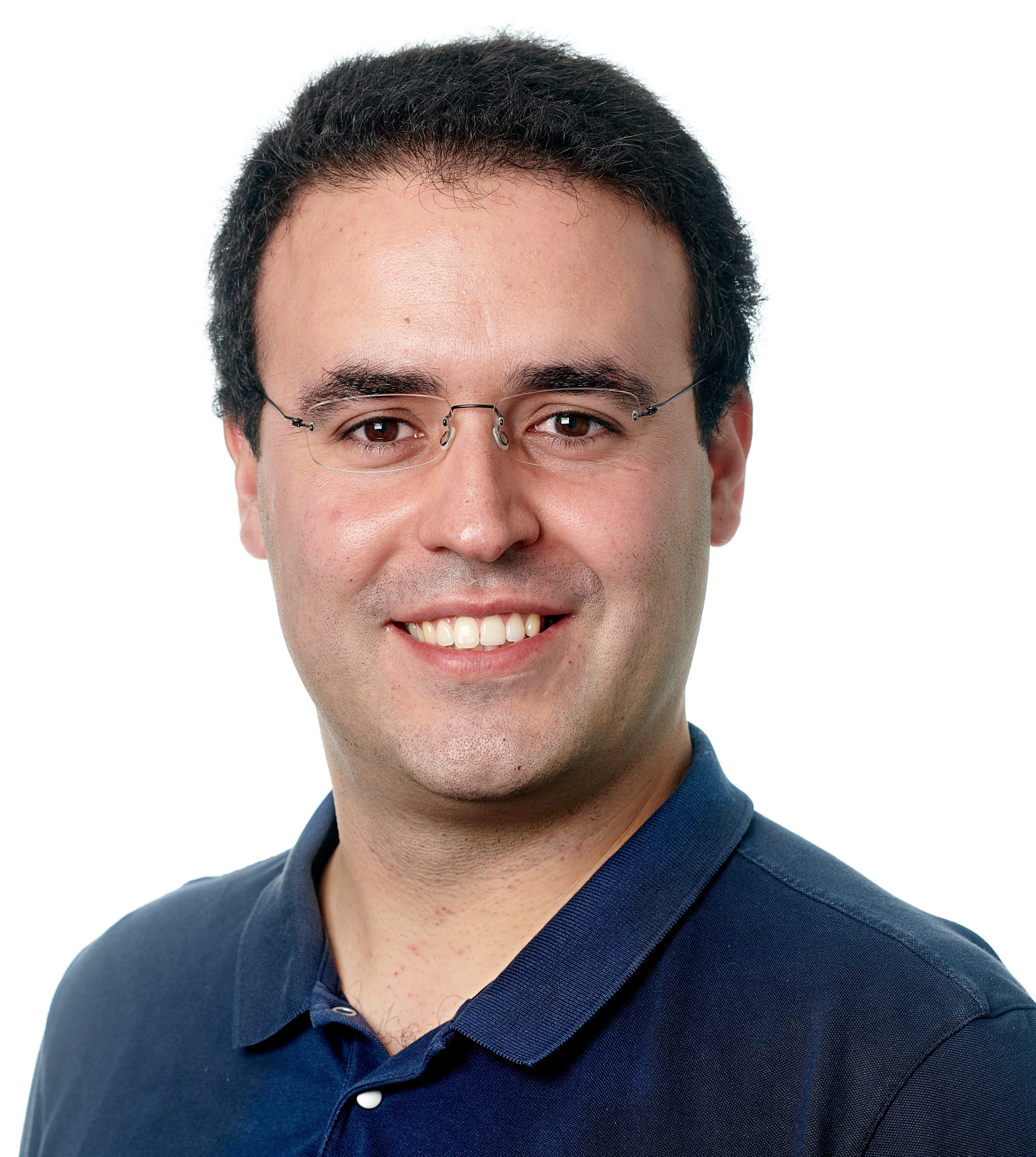}}]{Nuno K. Pratas} received the Licenciatura and M.Sc. degrees in electrical engineering from Instituto Superior Tecnico, Technical University of Lisbon, Lisbon, Portugal, in 2005 and 2007, respectively, and the Ph.D. degree in wireless communications from Aalborg University, Aalborg, Denmark, in 2012. He is currently a senior research specialist at Nokia. His research interests include wireless communications, networks and development of analysis tools for communication systems for 5G and 6G communications systems, in particular for sidelink use cases.
\end{IEEEbiography}
\begin{IEEEbiography}[{\includegraphics[width=1in,height=1.25in,clip,keepaspectratio]{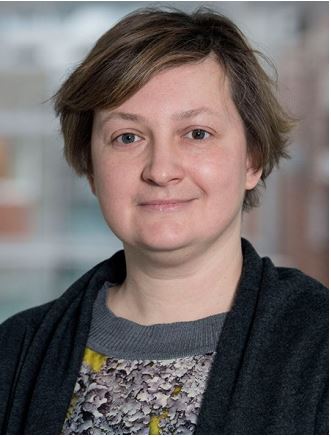}}]{Tatiana Kozlova Madsen} received Ph.D. in mathematics from Lomonosov Moscow State University, Russia in 2000. Since 2001 she is
working at Department of Electronic Systems, Aalborg University,
Denmark where she currently holds a position of an Associate Professor in Wireless Networking. Her research interests lie in the wide area
of wireless networking, including Quality of Service and performance
optimization of converging networks, communication protocols for IoT
systems and mesh networks and methods and tools for performance
evaluation of communication systems. She has been involved in a
number of national and international projects developing network
architectures, network protocols and solutions for Intelligent Transportation Systems; Smart Grids; automotive and train industries.
\end{IEEEbiography}
\begin{IEEEbiography}[{\includegraphics[width=1in,height=1.25in,clip,keepaspectratio]{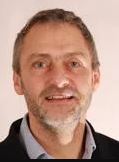}}]{Preben Mogensen} became a full professor at Aalborg University
	in 2000, where he is currently leading the Wireless Communication Networks Section. He is also a principal scientist in the
	Standardization \& Research Lab of Nokia Bell Labs. His current
	research interests include industrial use cases for 5G, 5G evolution, and 6G. He is a Bell Labs Fellow.
\end{IEEEbiography}
%%
%\begin{IEEEbiography}[{\includegraphics[width=1in,height=1.25in,clip,keepaspectratio]{a3.png}}]{Third C. Author, Jr.} (M'87) received the B.S. degree in mechanical 
%engineering from National Chung Cheng University, Chiayi, Taiwan, in 2004 
%and the M.S. degree in mechanical engineering from National Tsing Hua 
%University, Hsinchu, Taiwan, in 2006. He is currently pursuing the Ph.D. 
%degree in mechanical engineering at Texas A{\&}M University, College 
%Station, TX, USA.
%
%From 2008 to 2009, he was a Research Assistant with the Institute of 
%Physics, Academia Sinica, Tapei, Taiwan. His research interest includes the 
%development of surface processing and biological/medical treatment 
%techniques using nonthermal atmospheric pressure plasmas, fundamental study 
%of plasma sources, and fabrication of micro- or nanostructured surfaces. 
%
%Mr. Author's awards and honors include the Frew Fellowship (Australian 
%Academy of Science), the I. I. Rabi Prize (APS), the European Frequency and 
%Time Forum Award, the Carl Zeiss Research Award, the William F. Meggers 
%Award and the Adolph Lomb Medal (OSA).
%\end{IEEEbiography}

\EOD

\end{document}